\documentclass[format=acmsmall, review=false, screen=true]{acmart}

\usepackage{booktabs} 
\usepackage[ruled]{algorithm2e} 
\SetAlFnt{\small}
\SetAlCapFnt{\small}
\SetAlCapNameFnt{\small}
\SetAlCapHSkip{0pt}
\IncMargin{-\parindent}

\usepackage{longtable}
\usepackage{soul}
\usepackage{colortbl}
\usepackage{url}
\usepackage{subcaption}
\usepackage{caption}
\usepackage{booktabs}
\usepackage{tabularx}
\usepackage{algpseudocode}
\usepackage{amsfonts}
\usepackage{amsmath}
\usepackage{amssymb}
\usepackage{color}
\usepackage{xcolor}
\usepackage{verbatim}
\usepackage{booktabs}

\acmJournal{TWEB}
\acmVolume{0}
\acmNumber{0}
\acmArticle{0}
\acmYear{2018}
\acmMonth{0}
\copyrightyear{2018}

\setcopyright{acmlicensed}

\acmDOI{0000001.0000001}

\begin{document}
\title[Cashtag piggybacking: uncovering spam and bot activity in stock microblogs on Twitter]{Cashtag piggybacking: uncovering spam and bot activity in stock microblogs on Twitter}

\author{Stefano Cresci}
\orcid{0000-0003-0170-2445}
\affiliation{  \institution{Institute of Informatics and Telematics, IIT-CNR, Italy}
  \streetaddress{via G. Moruzzi, 1}
  \city{Pisa}
  \state{}
  \postcode{56124}
  \country{Italy}}
\email{stefano.cresci@iit.cnr.it}

\author{Fabrizio Lillo}
\orcid{}
\affiliation{  \institution{Department of Mathematics, University of Bologna, Italy}
  \streetaddress{}
  \city{}
  \state{}
  \postcode{}
  \country{Italy}}
\email{fabrizio.lillo@unibo.it}

\author{Daniele Regoli}
\orcid{}
\affiliation{  \institution{Scuola Normale Superiore of Pisa, Italy}
  \streetaddress{}
  \city{}
  \state{}
  \postcode{}
  \country{Italy}}
\email{daniele.regoli@sns.it}

\author{Serena Tardelli}
\orcid{}
\affiliation{  \institution{Institute of Informatics and Telematics, IIT-CNR, Italy}
  \streetaddress{}
  \city{}
  \state{}
  \postcode{}
  \country{Italy}}
\email{serena.tardelli@iit.cnr.it}

\author{Maurizio Tesconi}
\orcid{}
\affiliation{  \institution{Institute of Informatics and Telematics, IIT-CNR, Italy}
  \streetaddress{}
  \city{}
  \state{}
  \postcode{}
  \country{Italy}}
\email{maurizio.tesconi@iit.cnr.it}

\begin{abstract}
Microblogs are increasingly exploited for predicting prices and traded volumes of stocks in financial markets. However, it has been demonstrated that much of the content shared in microblogging platforms is created and publicized by bots and spammers. Yet, the presence (or lack thereof) and the impact of fake stock microblogs has never systematically been investigated before. Here, we study 9M tweets related to stocks of the 5 main financial markets in the US. By comparing tweets with financial data from Google Finance, we highlight important characteristics of Twitter stock microblogs. More importantly, we uncover a malicious practice -- referred to as \textit{cashtag piggybacking} -- perpetrated by coordinated groups of bots and likely aimed at promoting low-value stocks by exploiting the popularity of high-value ones. Among the findings of our study is that as much as 71\% of the authors of suspicious financial tweets are classified as bots by a state-of-the-art spambot detection algorithm. Furthermore, 37\% of them were suspended by Twitter a few months after our investigation. Our results call for the adoption of spam and bot detection techniques in all studies and applications that exploit user-generated content for predicting the stock market.
\end{abstract}




\maketitle

\renewcommand{\shortauthors}{S. Cresci et al.}

\makeatletter{}
\section{Introduction}
\label{sec:intro}
The exploitation of user-generated content in microblogs for the prediction of real-world phenomena, has recently gained huge momentum~\cite{schoen2013power}. An important application domain is that of finance, and in particular, stock market prediction. Indeed, a number of works developed algorithms and tools for extracting valuable information (e.g., sentiment scores) from microblogs and proved capable of predicting prices and traded volumes of stocks in financial markets~\cite{bollen2011twitter}. Notably, finance is increasingly relying on this information through the development of automatic trading systems.

All such works ground on the assumption that microblogs collectively represent a reliable proxy for the opinions of masses of users. Meanwhile however, evidence of fake accounts as well as spam and automated (bot) activities in social platforms is being reported at a growing rate~\cite{cresci2015fakefollowers,ferrara2016detection}. The existence of fictitious, synthetic content appears to be pervasive since it has been witnessed both in online discussions about important societal topics (e.g., politics, terrorism, immigration), as well as in discussions about seemingly less relevant topics, such as products on sale on e-commerce platforms, and mobile applications~\cite{cresci2017paradigm}. For instance, regarding politics, it has been demonstrated that bots tampered with recent US~\cite{bessi2016social}, Italian~\cite{cresci2016dna}, French~\cite{ferrara2017disinformation}, Japanese~\cite{schafer2017japan}, and -- to a minor extent -- German~\cite{Kupferschmidt1081,brachten2017strategies} political elections, as well as with online discussions about the 2016 UK Brexit referendum~\cite{bastos2017brexit}.

Thus, on the one hand, user-generated content in microblogs is being exploited for predicting trends in the stock market. On the other hand, without a thorough investigation, we run the risk that much of the content we rely on, is actually fake and possibly purposely created to mislead algorithms and users alike~\cite{papadopoulos2016overview}. Should this risk materialize, real-world consequences would be severe, as already anticipated by a few noteworthy events~\cite{ferrara2015manipulation}.
On May 6 2010, the Dow Jones Industrial Average had the biggest one-day drop in history, later called the \textit{Flash Crash}. After five months, an investigation concluded that one of the possible causes was an automated high-frequency trading system that had incorrectly assessed some information collected from the Web~\cite{hwang2012socialbots}. In 2013, the US International Press Officer's Twitter account got hacked and a false rumor was posted reporting that President Obama got injured during a terrorist attack. The fake news rapidly caused a stock market collapse that burned \$136B\footnote{\url{http://www.telegraph.co.uk/finance/markets/10013768/Bogus-AP-tweet-about-explosion-at-the-White-House-wipes-billions-off-US-markets.html}}. Then, in 2014, the unknown \textit{Cynk Technology} briefly became a \$6B worth company. Automatic trading algorithms detected a fake social discussion and begun to invest heavily in the company's shares. By the time analysts noticed the orchestration, investments had already turned into heavy losses\footnote{\url{http://mashable.com/2014/07/10/cynk/\#HD9o6llp6gqw}}.

\subsection{Contributions}
In a recent investigation~\cite{cresci2018fake}, we reported the first preliminary evidence of the presence of financial spam in stock microblogs, raising serious concerns over the reliability of such information. Here, we deepen our previous analyses by performing a number of additional experiments on co-occurring cashtags, on financial markets, and on suspicious users.
Specifically, we extend our previous work with the following novel and unpublished contributions:
\begin{itemize}
    \item we analyze co-occurring cashtags in financial tweets by focusing on their industrial and economic classification. In detail, we show that co-occurrences of stocks in suspicious tweets are not motivated by the fact that those stocks belong to the same industrial or economic sectors (\S~\ref{sec:spam-TRBC});
    \item since real-world relatedness (as expressed by industrial classification) is not a plausible explanation for co-occurring stocks, we then turn our attention to market capitalization. We demonstrate that, in suspicious tweets, high capitalization companies co-occur with low capitalization ones. Moreover, we show that this large difference can not be explained by the intrinsic characteristics of our dataset, but it is rather the consequence of an external action (\S~\ref{sec:spam-market-cap});
    \item we compare the social and financial importance of investigated companies, highlighting that stocks of one specific market (\texttt{OTCMKTS}) feature a suspiciously high social importance despite their low financial importance. This result is in contrast with measurements obtained for stocks of the other markets -- e.g., \texttt{NASDAQ}, \texttt{NYSE}, \texttt{NYSEARCA}, and \texttt{NYSEMKT} (\S~\ref{sec:spam-importance});
    \item we employ a state-of-the-art spambot detection technique to analyze authors of suspicious tweets. Results show that 71\% of suspicious users are classified as bots. Furthermore, 37\% of them also got suspended by Twitter a few months after our investigation (\S~\ref{sec:users}).
\end{itemize}

Summarizing, this study moves in the direction of investigating the presence of spam and bot activity in stock microblogs, thus paving the way for the development of intelligent financial-spam filtering techniques.
To reach our goal, we first collect a rich dataset comprising 9M tweets posted between May and September 2017, discussing stocks of the 5 main financial markets in the US. We enrich our dataset by collecting financial information from Google Finance about the 30,032 companies mentioned in our tweets. Cross-checking discussion patterns on Twitter against official data from Google Finance uncovers anomalies in tweets related to some low-value companies. Further investigation of this issue reveals a large-scale speculative campaign -- which we refer to as \textit{cashtag piggybacking} --  perpetrated by coordinated groups of bots and aimed at promoting low-value stocks by exploiting the popularity of high-value ones. Finally, we analyze a small subset of authors of suspicious tweets with state-of-the-art bot detection techniques, identifying 71\% ($18,509$ accounts) of them as bots.

\subsection{Cashtag piggybacking}
\label{sec:piggyback}

Results of our study uncover a large presence of bot accounts in stock microblogs on Twitter. More specifically, we thoroughly document a practice aimed at promoting low-cap stocks (mainly \texttt{OTCMKTS} stocks) by exploiting the popularity of high-cap ones.

We name this novel kind of spam as \textit{cashtag piggybacking}, by borrowing the concept of \textit{piggyback}\footnote{\url{https://en.wikipedia.org/wiki/Piggybacking_(data_transmission)}} from the field of computer networks~\cite{tanenbaum2014computer}. In many network protocols, a sender node is allowed to deliver short messages (e.g., ACKs to previous packets) to a receiver node, without sending a dedicated packet. In fact, the sender can postpone the short message until a new packet must be sent. At this time, the sender piggybacks (i.e., adds) the message as part of the outgoing packet. In network protocols, piggybacking allows to increase the efficiency in communications~\cite{tanenbaum2014computer}. Indeed, fewer packets are sent, since small amounts of information can be sent ``on top of the shoulders'' of other packets.

Within the context of stock microblogs, we show that coordinated groups of bots piggyback some low-value stocks ``on top of the shoulders'' of other high-value stocks. Hence, the \textit{cashtag piggybacking} name.

\subsection{Roadmap}
The remainder of this paper is organized as follows. Section~\ref{sec:relwork} discusses relevant related work in stock market prediction from social media, and in spam and bot detection. Then, Section~\ref{sec:dataset} describes the dataset used in this study. In Section~\ref{sec:peaks} we briefly provide an overview of the characteristics of our dataset and we describe the methodology adopted to identify suspicious tweets. In Section~\ref{sec:spam} we analyze suspicious tweets and financial markets from several viewpoints. Instead, in Section~\ref{sec:users} we turn our attention to the authors of the suspicious tweets, looking for bots among them. Section~\ref{sec:piggyback} gives the motivations and definition of the newly identified \textit{cashtag piggybacking} spam campaign. Section~\ref{sec:discussion} provides a critical discussion of our results, and finally, Section~\ref{sec:conc} draws conclusions and highlights some promising directions for future research and experimentation.

\makeatletter{}
\section{Related work}
\label{sec:relwork}
Since no study has previously addressed bot activity in stock microblogs, this section is organized so as to separately survey previous work either related to the exploitation of user-generated content for financial purposes, or to spam and bot characterization.

\subsection{Finance and social media}
\label{sec:soa-finance}
Works in this field are based on the idea underlying the Hong-Page theorem~\cite{hong2004groups}. Such theorem, when cast in the financial domain, states that user-generated messages about a company's future prospects provide a rich and diverse source of information, in contrast to what the small number of traditional financial analysts can offer.

Starting from the general assumption of the Hong-Page theorem, much effort has been devoted towards the detection of correlations between metrics extracted from social media posts and stock market prices. In particular, \textit{sentiment} metrics have been widely used as a predictor for stock prices and other economic indicators~\cite{gilbert2010widespread,bollen2011modeling,sprenger2011tweettrader,chen2014wisdom,ranco2015effects,gabrovvsek2017twitter}. The primary role played by the sentiment of the users as a financial predictor is also testified by the interest in developing domain-specific sentiment classifiers for the financial domain~\cite{smailovic2014stream,cortis2017semeval}. Others have instead proposed to exploit the overall volume of tweets about a company~\cite{mao2012correlating} and the topology of stock networks~\cite{ruiz2012correlating} as predictors of financial performance. Specifically, authors of~\cite{mao2012correlating} envisioned the possibility to automatically buy or sell stocks based on the presence of a peak in the volume of tweets. However, subsequent work~\cite{zheludev2014can} evaluated the informativeness of sentiment- and volume-derived predictors, showing that the sentiment of tweets contains significantly more information for predicting stock prices than just their volume. The role of \textit{influencers} in social media has also been identified as a strong contributing factor to the formation of market trends~\cite{cazzoli2016large}.
Others have instead used weblogs for studying the relationships between different companies~\cite{kharratzadeh2012weblog}. In detail, co-occurrences of stock mentions in weblogs have been exploited to create a graph of companies, which was subsequently clustered. Authors have verified that companies belonging to the same clusters feature strong correlations in their stock prices. This methodology can be employed for market prediction and as a portfolio-selection method, which has been shown to outperform traditional strategies based on company sectors or historical stock prices.

Another line of research focused on the exploitation of social media content for monitoring and predicting firm equity value. As an example, the study in~\cite{yu2013impact} investigated the effect of social media and conventional media, their relative importance, and their inter-relatedness on short-term firm equity value prediction. Findings indicated that social media has a stronger relationship with firm equity value than conventional media, while social and conventional media have a strong interaction effect on stock performance. Similarly, in~\cite{luo2013social} authors focused on the effects of social media-derived metrics compared with conventional online company behavioral metrics. Results derived from autoregressive models suggested that social media-derived metrics (e.g., weblogs and consumer ratings) are significant leading indicators of firm equity value. Even more interestingly, conventional online behavioral metrics (Google searches and Web traffic) have a significant yet substantially weaker predictive relationship with firm equity value than social media metrics. Another study~\cite{luo2013consumer} from the same authors assessed the extent to which ``consumer buzz'', in the form of user-generated reviews, recommendations, and blog posts, influence firm value. Results support the dynamic relationships of buzz and Web traffic with firm value, and the related mediation effects of buzz and traffic. The study also uncovered significant market competition effects, including effects of both a firm's own and its rivals' buzz and traffic.

Nowadays, results of studies such as those briefly surveyed in this section are leveraged for the development of automatic trading systems that are largely fed with social media-derived information~\cite{feldman2013techniques}. As a consequence, such automatic systems can potentially suffer severe problems caused by large quantities of fictitious posts. As discussed in the next section, the presence of social bots -- and of the fake content they produce -- is so widespread as to represent a serious, tangible threat to these, and other, systems~\cite{gilani2017bots}.

\subsection{Spam and bots in social media}
\label{sec:soa-spambots}
Since our study is aimed at verifying the presence and the impact of spam and bot activity in stock microblogs, in this section we focus on discussing previous work about the characterization and detection of spam and bots in social media.

Many developers of spammer accounts make use of bots in order to simultaneously and continuously post a great deal of spam content. This is one of the reasons why, despite bots being in rather small numbers when compared to legitimate users, they nonetheless have a profound impact on content popularity and activity in social media~\cite{aiello2012people,gilani2017bots}. In addition, bots are driven so as to act in a \textit{coordinated} and \textit{synchronized} way, thus amplifying their effects~\cite{ratkiewicz2011detecting,zhang2016detecting}. Another problem with bots is that they \textit{evolve} over time, in order to evade established detection techniques~\cite{yang2013,cresci2018proaction}. Hence, newer bots often feature advanced characteristics that make them way harder to detect with respect to older ones. Recently, a general-purpose overview of the landscape of automated accounts was presented in~\cite{ferrara2016rise}. This work testifies the emergence of a new wave of social bots, capable of mimicking human behavior and interaction patterns in social media better than ever before. A subsequent study~\cite{cresci2017paradigm} compared ``traditional'' and ``evolved'' bots in Twitter, and demonstrated that the latter are almost completely undetected by platform administrators. Moreover, a crowdsourcing campaign showed that even tech savvy users are incapable of accurately identifying the evolved bots.

Since bots and spammers evolved, putting in place complex mechanisms to evade existing detection systems, scholars and platform administrators tried to keep pace by proposing powerful techniques based on profile~\cite{cresci2015fakefollowers,lee2014early,alowibdi2014detecting}, posting~\cite{wu2015social,badri2016uncovering,lee2013warningbird}, and network~\cite{yu2010sybillimit,ghosh2012understanding,akoglu2010oddball,jiang2014inferring} 
characteristics of the accounts.
The study presented in~\cite{cresci2017paradigm} however demonstrated that also the majority of these bot detection techniques, which are based on off-the-shelf machine learning algorithms applied for analyzing of one account at a time, are unable to effectively detect the evolved bots. In order to overcome this limitation a recent stream of research proposed ad-hoc detection techniques for the collective analysis of groups of accounts, rather than single accounts~\cite{jiang2016inferring,jiang2016catching,yu2015glad,jiang2016spotting,cresci2016dna,cresci2017tdsc}. These techniques achieved better detection results than previous ones~\cite{cresci2017paradigm}, and represent nowadays the last bulwark against pervasive malicious accounts in social media.

However, the battle is far from over. Indeed, given this worrying picture, it is not surprising that bots have recently proven capable of influencing the public opinion for many crucial topics~\cite{bessi2016social,ferrara2017disinformation,bastos2017brexit} and in many different ways, such as by spreading fake news~\cite{shao2016hoaxy} or by artificially inflating the popularity of certain posts~\cite{beutel2013copycatch} and public characters~\cite{cresci2015fakefollowers}. 
The combination of automatic trading systems feeding on social media data and the pervasive presence of spam and bots, motivates our investigation on the presence of spam and bots in stock microblogs.
Moreover, the financial domain has already been proven to have peculiar characteristics with respect to many information processing tasks (e.g., ranking~\cite{ceccarelli2016ranking} and filtering~\cite{tang2017echo} content, expert finding~\cite{wang2017value}, etc.) so as to require ad-hoc analyses, such as the one carried out in this work.

\makeatletter{}
\section{Dataset}
\label{sec:dataset}

Our dataset for this study is composed of: (i) stock microblogs collected from Twitter, and (ii) financial information collected from Google Finance.

\subsection{Twitter data collection}
\label{sec:dataset-twitter}
\begin{figure*}[t]
    \centering
    \begin{subfigure}[t]{.40\textwidth}
        \includegraphics[width=\textwidth]{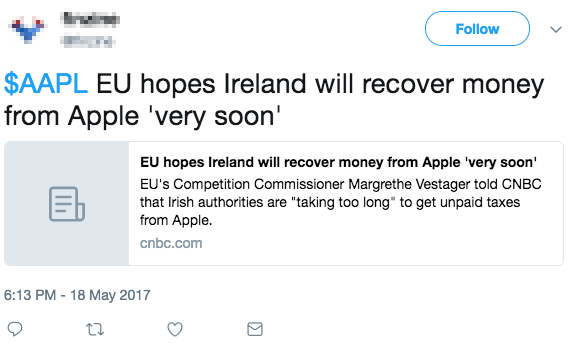}
    \end{subfigure}\hspace{.05\textwidth}    \begin{subfigure}[t]{.40\textwidth}
        \includegraphics[width=\textwidth]{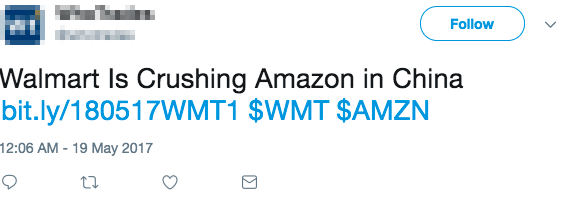}
    \end{subfigure}    \caption{Sample tweets with the \texttt{\$AAPL}, \texttt{\$WMT}, and \texttt{\$AMZN} cashtags.\label{fig:tweet-AAPL}}
\end{figure*}

Twitter users follow the convention of tagging stock microblogs with so-called \textit{cashtags}. The cashtag of a company is composed of a dollar sign followed by its ticker symbol (e.g., \texttt{\$AAPL} is the cashtag of \textit{Apple, Inc.}). Figure~\ref{fig:tweet-AAPL} shows two sample tweets with the \texttt{\$AAPL}, \texttt{\$WMT}, and \texttt{\$AMZN} cashtags. Similarly to hashtags, cashtags are visually highlighted on Twitter's interface can be used as an efficient mean to filter content and to collect data about given companies~\cite{hentschel2014follow}. For this reason, we based our Twitter data collection on an official list of cashtags. Specifically, we first downloaded a list of 6,689 stocks traded on the most important US markets (e.g., \texttt{NASDAQ}, \texttt{NYSE}) from the official \texttt{NASDAQ} Web site\footnote{\url{http://www.nasdaq.com/screening/company-list.aspx}}. Then, we collected all tweets shared between May and September 2017, containing at least one cashtag from the list. Data collection from Twitter has been carried out by exploiting Twitter's Streaming APIs\footnote{\url{https://developer.twitter.com/en/docs/tweets/filter-realtime/overview}}.
After our 5 months data collection, we ended up with $\sim$9M tweets (of which 22\% are retweets), posted by $\sim$2.5M distinct users, as shown in Table~\ref{tab:dataset}.

As a consequence of our data collection strategy, every tweet in our dataset contains at least one cashtag from the starting list. However, many collected tweets contain more than one cahstag, many of which are related to companies not included in our starting list. Indeed, overall we collected data about 30,032 companies traded across 5 different markets.

\makeatletter{}\begin{table*}[t]
	\footnotesize
	\centering
	\begin{tabular}{lcrrrcrrr}
	   
		\toprule
		&& \multicolumn{3}{c}{\textbf{financial data}} && \multicolumn{3}{c}{\textbf{twitter data}}\\
		\cmidrule{3-5} \cmidrule{7-9}
		\textbf{markets} && \textit{companies} & \textit{median cap. (\$)} & \textit{total cap. (\$B)} && \textit{users} & \textit{tweets} & \textit{retweets (\%)}\\
		\midrule
		\texttt{NASDAQ}	    && 3,013	& 365,780,000   & 10,521    && 252,587			& 4,017,158			& 1,017,138 (25\%)		\\
		\texttt{NYSE}	    && 2,997	& 1,810,000,000 & 28,692	&& 265,618			& 4,410,201			& 923,123 (21\%)		\\
		\texttt{NYSEARCA}	&& 726	    & 245,375,000   & 2,227     && 56,101			& 298,445			& 157,101 (53\%)		\\
		\texttt{NYSEMKT}	&& 340	    & 78,705,000    & 256	    && 22,614			& 196,545			& 63,944 (33\%)		\\
		\texttt{OTCMKTS}	&& 22,956	& 31,480,000    & 45,457	&& 64,628	        & 584,169			& 446,293 (76\%)		\\
		\midrule
		\textbf{total}	    && 30,032	& --	            & 87,152     && 467,241			    & 7,855,518			        & 1,802,705	(23\%)	\\
		\bottomrule
	\end{tabular}
	\caption{Financial and social dataset composition. Total values of \textit{users}, \textit{tweets}, and \textit{retweets} only count distinct items and thus do not equal to the sum of previous rows.}
	\label{tab:dataset}
\end{table*}

\subsection{Financial data collection}
\label{sec:dataset-financial}
We enriched our Twitter dataset by collecting financial information about each of the 30,032 companies found in our tweets. Financial information have been collected from public company data hosted on Google Finance\footnote{\url{https://www.google.com/finance}}. Among collected financial information, is the \textit{market capitalization} (market cap) of a company and its \textit{industrial classification}.

The capitalization is the total dollar market value of a company.
For a given company $i$, it is computed as the share price $P(s_i)$ times the number of outstanding shares $|s_i|$: $C_i = P(s_i) \times |s_i|$.
In our study, we take the market cap of a company into account, since it allows us to compare the financial value of that company with its social media popularity and engagement\footnote{In the remainder, share prices and market capitalizations are considered as of July 4, 2017.}. In Table~\ref{tab:dataset} we report the median capitalization of the companies for each considered market.
As shown, important markets such as \texttt{NYSE} and \texttt{NASDAQ} trade, on average, stocks with higher capitalization than those traded in minor markets.

\begin{figure}[t]
\centering
\includegraphics[width=0.7\textwidth]{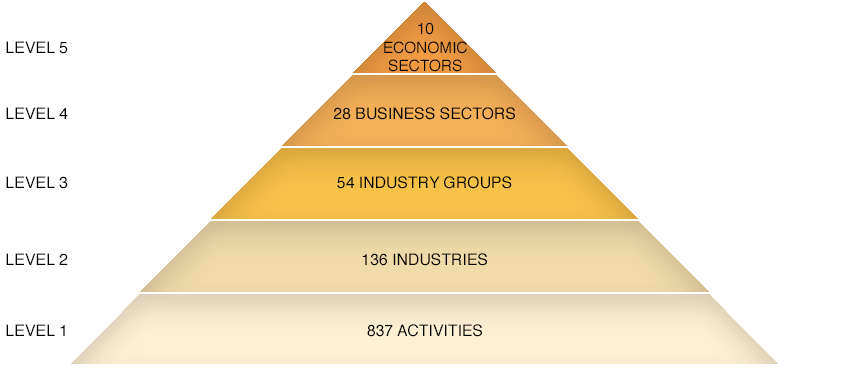}
\caption{Thomson Reuters Business Classification (TRBC) hierarchical schema.\label{fig:TRBC-classification}}
\end{figure}

Industrial classification is expressed via the Thomson Reuters Business Classification\footnote{\url{https://financial.thomsonreuters.com/en/products/data-analytics/market-data/indices/trbc-indices.html}} (TRBC). As shown in Figure~\ref{fig:TRBC-classification}, TRBC is a 5-level hierarchical sector and industry classification, widely used in the financial domain for computing sector-specific indices.
At the topmost (coarse-grained) level TRBC classifies companies into 10 economic sectors, while at the lowest (fine-grained) level companies are divided into 837 different activities. A few examples of the TRBC industrial classification are reported in Table~\ref{tab:TRBC-classification-examples}. In our study, we compare companies belonging to the same category, across all 5 levels of TRBC.

\makeatletter{}\begin{table*}[t]
	\scriptsize
	\centering
	\begin{tabular}{lcp{0.14\textwidth}p{0.15\textwidth}p{0.125\textwidth}p{0.125\textwidth}p{0.11\textwidth}}
		\toprule
		&& \multicolumn{5}{c}{\textbf{TRBC levels}} \\
		\cmidrule{3-7}
		\textbf{ticker} & \textbf{company} & \textit{activity} & \textit{industry} & \textit{industrial group} & \textit{business sector} & \textit{economic sector}  \\
		\midrule
		\texttt{AAPL}	    & Apple, Inc.	& Computer Hardware-NEC	    & Computer Hardware			& Computers, Phones \& Household Electronics			& Technology Equipment	    & Technology	    \\ [25pt]
		\texttt{GOOG}	    & Alphabet, Inc.	& Search Engines		& Internet Services			& Software \& IT Services			& Software \& IT Services		& Technology     \\ [15pt]
		\texttt{JNJ}	& Johnson \& Johnson	    & Pharmaceutic-NEC		& Pharmaceutic			& Pharmaceutic			& Pharmaceutics \& Medical Research		& Healthcare     \\
		\bottomrule
	\end{tabular}
	\caption{Examples of TRBC classifications.}
	\label{tab:TRBC-classification-examples}
\end{table*}

\makeatletter{}

\section{Analysis of stock microblogs}

\label{sec:peaks}
\begin{figure*}[t]
    \centering
    \begin{subfigure}[t]{.9\textwidth}
        \includegraphics[width=\textwidth]{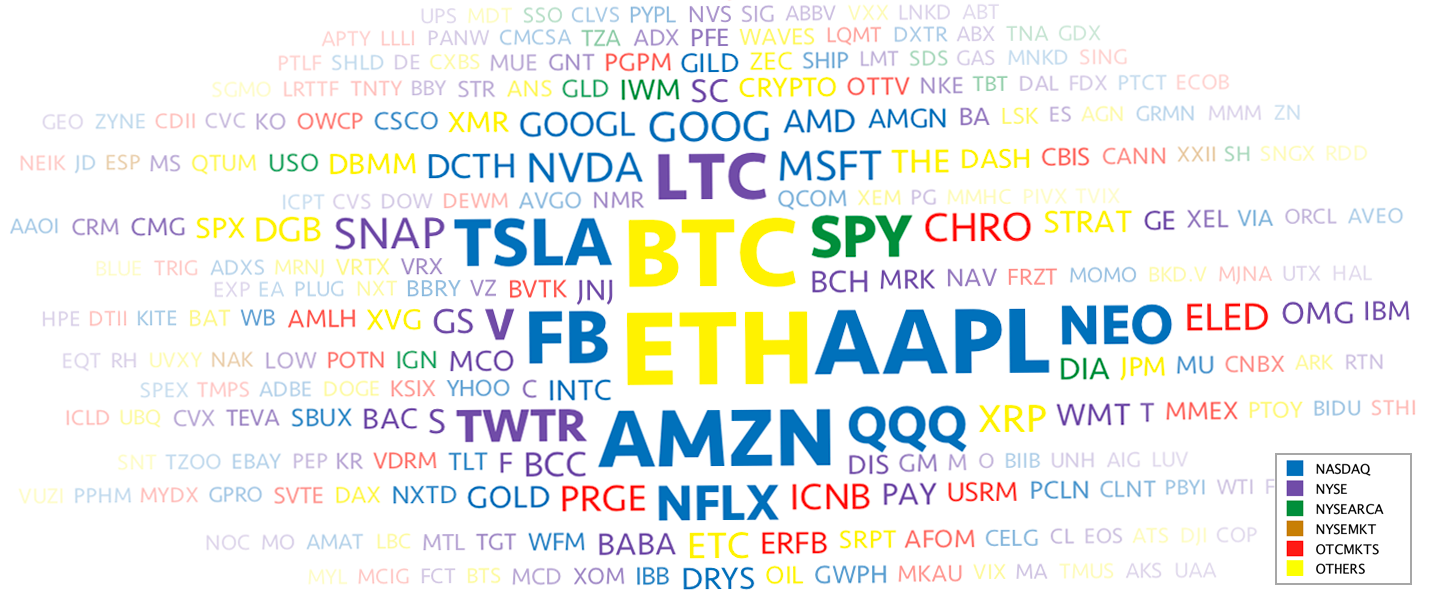}
        \caption{Cashtag-cloud of most tweeted companies.\label{fig:cashtags-most-tweeted}}
    \end{subfigure}\\%
    \begin{subfigure}[t]{.35\textwidth}
        \includegraphics[width=\textwidth]{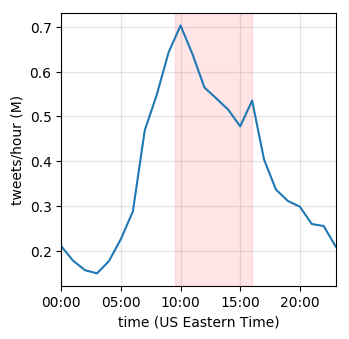}
        \caption{Mean tweet volume per hour. Peak hours overlap with the opening hours of the New York Stock Exchange (red band).\label{fig:tweets-per-hour}}
    \end{subfigure}\hspace{.15\textwidth}    \begin{subfigure}[t]{.35\textwidth}
        \includegraphics[width=\textwidth]{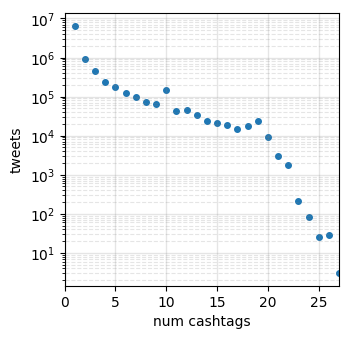}
        \caption{Distribution of the number of cashtags per tweet.\label{fig:cashtags-per-tweet}}
    \end{subfigure}    \caption{Overall statistics about our dataset.\label{fig:data-overview}}
\end{figure*}

\begin{figure*}[t]
    \centering
    \includegraphics[width=0.95\textwidth]{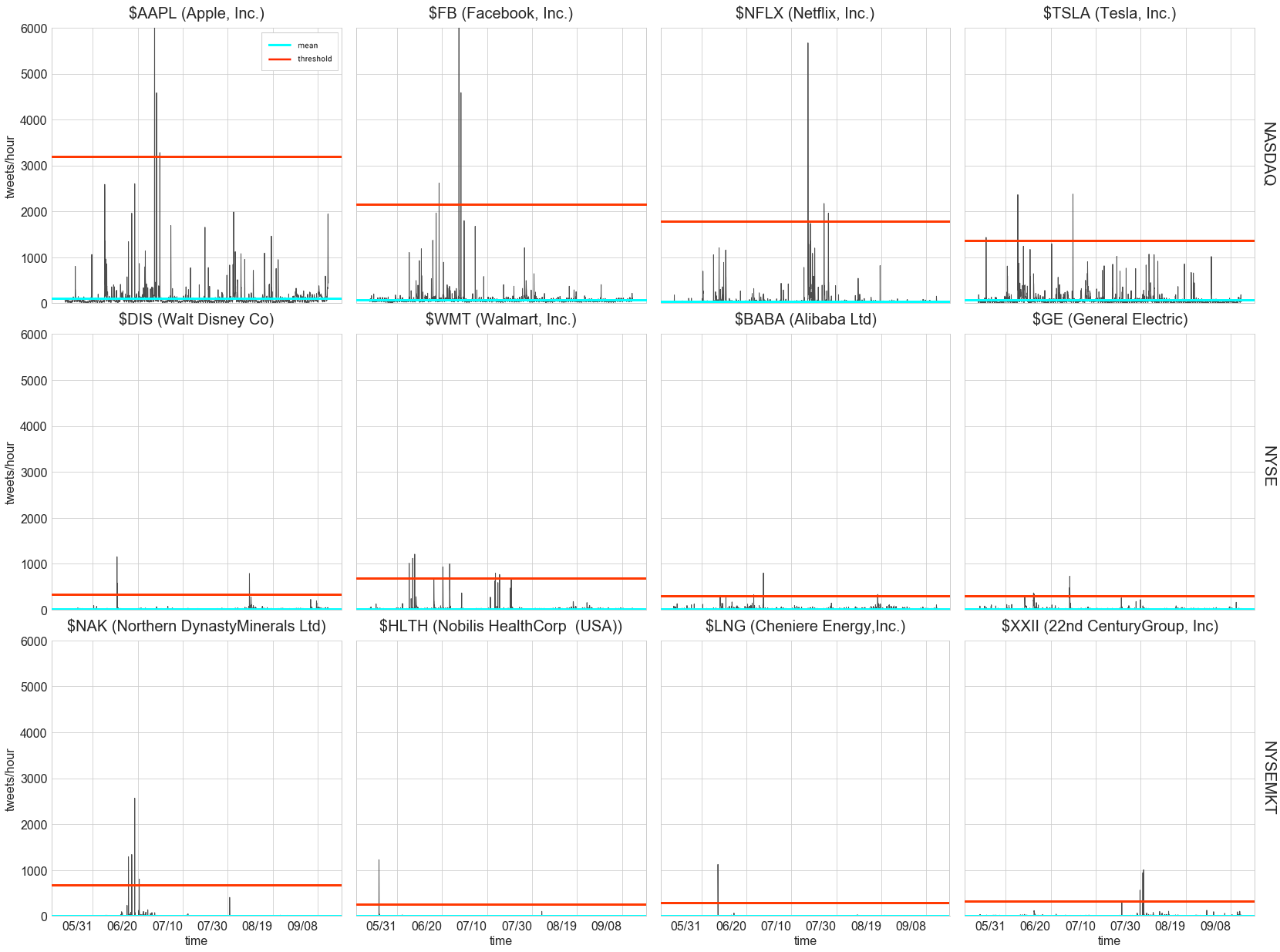}
    \caption{Examples of stock time series, for 12 highly tweeted stocks. Mean values are marked with cyan solid lines and thresholds above which peaks are detected ($K = 10$) are marked with red solid lines.\label{fig:time-series}}
\end{figure*}

\subsection{Dataset overview}
\label{sec:data-overview}
Surprisingly, the vast majority (76\%) of companies mentioned in our dataset do not belong to the \texttt{NASDAQ} list and are traded in \texttt{OTCMKTS}, as shown in Table~\ref{tab:dataset}.
Having so many \texttt{OTCMKTS} companies in our dataset is already an interesting finding, considering that our data collection grounded on a list of high-capitalization (high-cap) companies.
\texttt{OTCMKTS} is a US financial market for over-the-counter transactions, and thus it has far less stringent requirements than those needed from \texttt{NASDAQ}, \texttt{NYSE}, \texttt{NYSEARCA}, and \texttt{NYSEMKT}. For this reason, many small companies opt to be traded in \texttt{OTCMKTS} instead of the more requiring markets. However, in addition to small-cap companies, \texttt{OTCMKTS} also trades \textit{American depositary receipts} (ADRs)\footnote{\url{https://en.wikipedia.org/wiki/American_depositary_receipt}}, which allow to trade in US markets the stocks of non-US companies, otherwise only traded in other foreign markets (e.g., stocks of \textit{Samsung Electronics Co., Ltd.} would only be traded in the Korea Exchange).
\texttt{OTCMKTS} also trades \textit{Convertible Preferred stocks}\footnote{\url{https://en.wikipedia.org/wiki/Preferred_stock}}, which are a particular kind of stocks that give more guarantees to investors with respect to common stocks. Other types of assets might be traded in this market. In our study, we do not discriminate between different types of assets traded in OTCMKTS and we rely on the financial information contained in Google Finance, irrespectively of the kind of traded stocks. 

Thus, from a company viewpoint, our dataset is dominated by stocks traded in \texttt{OTCMKTS}. However, \texttt{OTCMKTS} companies play a marginal role from both a financial and social viewpoint, having low median capitalization and small numbers of tweets, the vast majority of which are retweets. In contrast, companies from \texttt{NASDAQ} and \texttt{NYSE} have high capitalization and are mentioned in many tweets, with low percentage of retweets.

In the following, we report on some of the general characteristics of our dataset.
Figure~\ref{fig:cashtags-most-tweeted} shows a cashtag-cloud representing the most tweeted companies in our dataset. In figure, cashtags are color-coded so as to visually highlight companies traded in different markets. The most tweeted companies in our dataset are in line with recent trends (e.g., the \texttt{\$BTC} (\textit{Bitcoin}) and \texttt{\$ETH} (\textit{Ethereum}) cryptocurrencies) and with findings of previous works~\cite{hentschel2014follow,alonso2014kondenzer} (e.g., \texttt{\$AAPL} leading the way, followed by \texttt{\$AMZN}, \texttt{\$FB}, and \texttt{\$TSLA}). Notably, no company from \texttt{OTCMKTS} appears among top mentioned companies, but instead they play a rather marginal role.
Figure~\ref{fig:tweets-per-hour} shows the mean volume of tweets collected per hour. The largest surge of tweets occurs between 10am and 5pm (US Eastern time), which almost completely overlaps with the opening hours of the New York Stock Exchange (9:30am to 4pm). This fact further highlights the strong relation between stock microblogs and the real-world stock market.
Finally, as previously introduced, many stock microblogs contain more than one cashtag (e.g., the right-hand side tweet in Figure~\ref{fig:tweet-AAPL}). Figure~\ref{fig:cashtags-per-tweet} shows the distribution of distinct cashtags per tweet, with a mean value of 2 cashtags/tweet.

\subsection{Stock time series analysis}
\label{sec:stock-analysis}

In order to uncover possible malicious behaviors related to stock microblogs, we carry out a fine-grained analysis of our data. Specifically, we build and analyze the hourly time series of each of the 6,689 stocks downloaded from the \texttt{NASDAQ} Web site. Given a stock $i$, its time series is defined as $\mathbf{s}_i = (s_{i,1}, s_{i,2}, \dots, s_{i,N})$, with $s_{i,j}$ being the number of tweets that mentioned the stock $i$ during the hour $j$. Figure~\ref{fig:time-series} shows some examples of our stock time series, for 12 highly tweeted stocks across 3 markets (\texttt{NASDAQ}, \texttt{NYSE}, and \texttt{NYSEMKT}). As shown in figure, stock time series are characterized by long time spans over which tweet discussion volumes remain rather low, occasionally interspersed by large discussion spikes. This behavior is consistent with what has been previously observed in Twitter for other phenomena (e.g., communication patterns related to emergency events~\cite{avvenuti2017nowcasting}). Indeed, the \textit{bursty} an \textit{spiky} characteristics of social communications have been recently explained as a direct consequence of human dynamics~\cite{karsai2012universal}.

\begin{figure*}[t]
	\begin{minipage}[b]{0.375\textwidth}
        \centering
        \includegraphics[width=\textwidth]{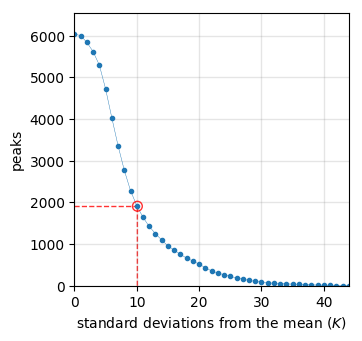}
        \caption{Number of peaks detected, as a function of $K$.\label{fig:peaks-as-function-of-k}}
	\end{minipage}
	\hspace{0.05\textwidth}
	\begin{minipage}[b]{0.55\textwidth}
        \captionsetup[subfigure]{labelformat=empty}
        \centering
        \begin{subfigure}{.45\textwidth}
                        \includegraphics[width=\textwidth]{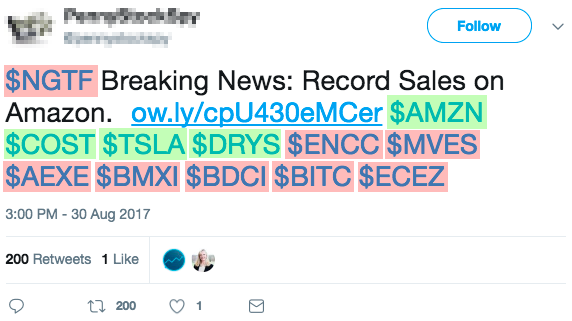}
        \end{subfigure}\hspace{.02\textwidth}        \begin{subfigure}{.45\textwidth}
                        \includegraphics[width=\textwidth]{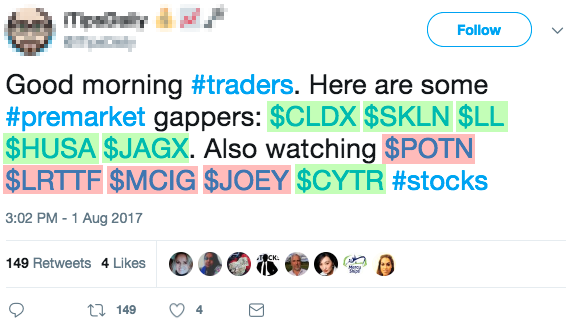}
        \end{subfigure}\hspace{.02\textwidth}        \begin{subfigure}{.45\textwidth}
                        \includegraphics[width=\textwidth]{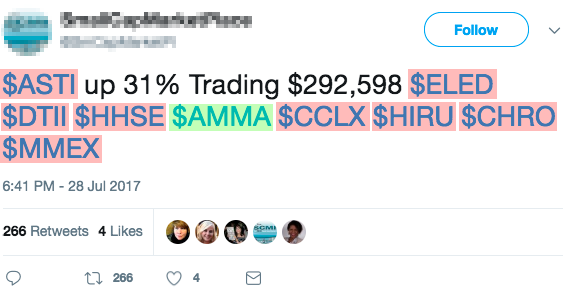}
        \end{subfigure}\hspace{.02\textwidth}        \begin{subfigure}{.45\textwidth}
                        \includegraphics[width=\textwidth]{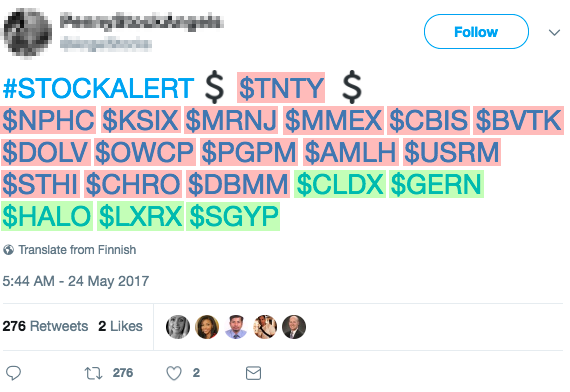}
        \end{subfigure}
        \caption{Examples of suspicious peak tweets. In every tweet, a few cashtags of high-cap stocks (green-colored) co-occur with many cashtags of low-cap stocks (red-colored).\label{fig:anomalous-tweets}}
	\end{minipage}
\end{figure*}

To give a better characterization of this phenomenon we ran a simple anomaly detection technique on all the 6,689 time series. As typically done in many time series analysis tasks, our anomaly detection technique is designed so as to detect a peak $p_{i,j}$ in a time series $\mathbf{s}_i$ \textit{iff} the tweet volume for the hour $j$ deviates from the mean tweet volume $\bar{\mathbf{s}}_i$ by a number $K$ of standard deviations:
\[
p_{i,j} \iff s_{i,j} > \bar{\mathbf{s}}_i + K \times \sigma(\mathbf{s}_i)
\]
The parameter $K$ determines the number of peaks found by our anomaly detection technique. In fact, a bigger $K$ implies that a larger deviation from the mean is needed in order to detect a peak. Figure~\ref{fig:peaks-as-function-of-k} shows the number of peaks detected in our time series, as a function of the parameter $K$. For the remainder of our analysis we set $K = 10$, which represents a trade-off between the height of considered peaks and the number of peaks to analyze. This choice of $K$ results in 1,926 peaks detected in our time series. Time series depicted in Figure~\ref{fig:time-series} also show mean values (cyan solid line) and the $10\sigma$ threshold (red solid line) above which peaks are detected.

Next, we are interested in analyzing the tweets that generated the peaks (henceforth, \textit{peak tweets}). In detail, a peak $p_{i,j}$ is composed of a set of tweets $\mathbf{t}_{i,j}$, such that each tweet $t \in \mathbf{t}_{i,j}$ contains the cashtag related to the stock $i$ and has been posted during the hour $j$ (i.e., the peak hour):
\[
\mathbf{t}_{i,j} = \{t^1_{i,j}, t^2_{i,j}, \dots, t^M_{i,j}\}, \quad M = s_{i,j}
\]
Thus, for each of the 1,926 peaks $p_{i,j}$ we analyze the corresponding set of tweets $\mathbf{t}_{i,j}$. We find out that, on average, 60\% of tweets $t \in \mathbf{t}$ are retweets. In other words, the peaks identified by our anomaly detection technique are largely composed of retweets. In addition, considering that our time series have hourly granularity, those retweets also occurred within a rather limited time span, in a \textit{bursty} fashion. This finding is particularly interesting also considering that in all our dataset, we had only 23\% retweets, versus 60\% measured for peak tweets.

\begin{figure*}[t]
    \centering
    \begin{subfigure}[t]{.35\textwidth}
        \includegraphics[width=\textwidth]{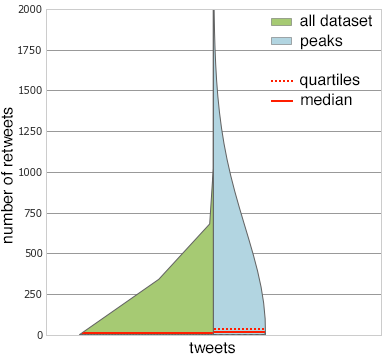}
        \caption{Retweets per tweet.\label{fig:beanplot-retweets}}
    \end{subfigure}\hspace{.1\textwidth}    \begin{subfigure}[t]{.35\textwidth}
        \includegraphics[width=\textwidth]{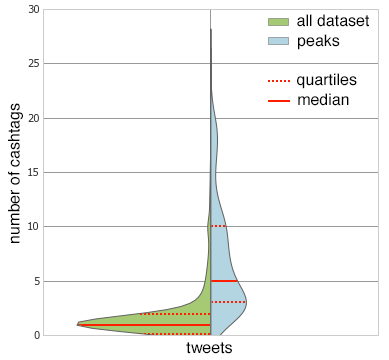}
        \caption{Cashtags per tweet.\label{fig:beanplot-cashtags}}
    \end{subfigure}    \caption{Beanplots showing the differences in the number of retweets per tweet and in the number of cashtags per tweet, for all tweets of the dataset (green-colored) and for peak tweets (light blue-colored). Peak tweets feature a higher number of retweets and a higher number of cashtags per tweet.\label{fig:beanplots}}
\end{figure*}

We also analyzed tweets $t \in \mathbf{t}$ by considering the co-occurrences of stocks. From this analysis we see that tweets $t \in \mathbf{t}$ typically contain many more cashtags than tweets $t \notin \mathbf{t}$. The cashtags that co-occur in peak tweets seem unrelated, and the authors of those tweets don't provide further information to explain such co-occurrences. As an example, Figure~\ref{fig:anomalous-tweets} shows 4 of such suspicious tweets. In figure, in every tweet, a few cashtags of high-capitalization (high-cap) stocks co-occur with many cashtags of low-cap stocks. The distributions of the number of retweets per tweet, and of the number of cashtags per tweet, are shown in figures~\ref{fig:beanplot-retweets} and~\ref{fig:beanplot-cashtags} respectively. In figure, the distributions are shown with beanplots and allow to compare values measured for the whole dataset (green-colored), with those measured only in peak tweets (light blue-colored).

The characteristics of peak tweets previously highlighted -- that is, the percentage of retweets and the number of co-occurring cashtags -- differ significantly from those measured for the whole dataset. The reason for this peculiar phenomenon could be related to some real-world news or event, that motivates the surge of retweets and the co-occurrences of different cashtags. However, such differences could also be the consequence of a shady, malicious activity. Indeed, there have already been reports of large groups of bots that coordinately and simultaneously alter popularity and engagement metrics of Twitter users and content~\cite{beutel2013copycatch,ferrara2016detection}. In particular, mass retweets have been identified as one mean to artificially increase the popularity of certain content~\cite{cresci2017paradigm}.

\makeatletter{}
\section{Anatomy of financial spam}
\label{sec:spam}
In this section we evaluate different hypotheses in order to thoroughly understand the reasons why so many seemingly-unrelated cashtags co-occur in peak tweets, and the reason for the high percentage of retweets in peaks.

\begin{figure*}[t]
    \centering
    \includegraphics[width=0.8\textwidth]{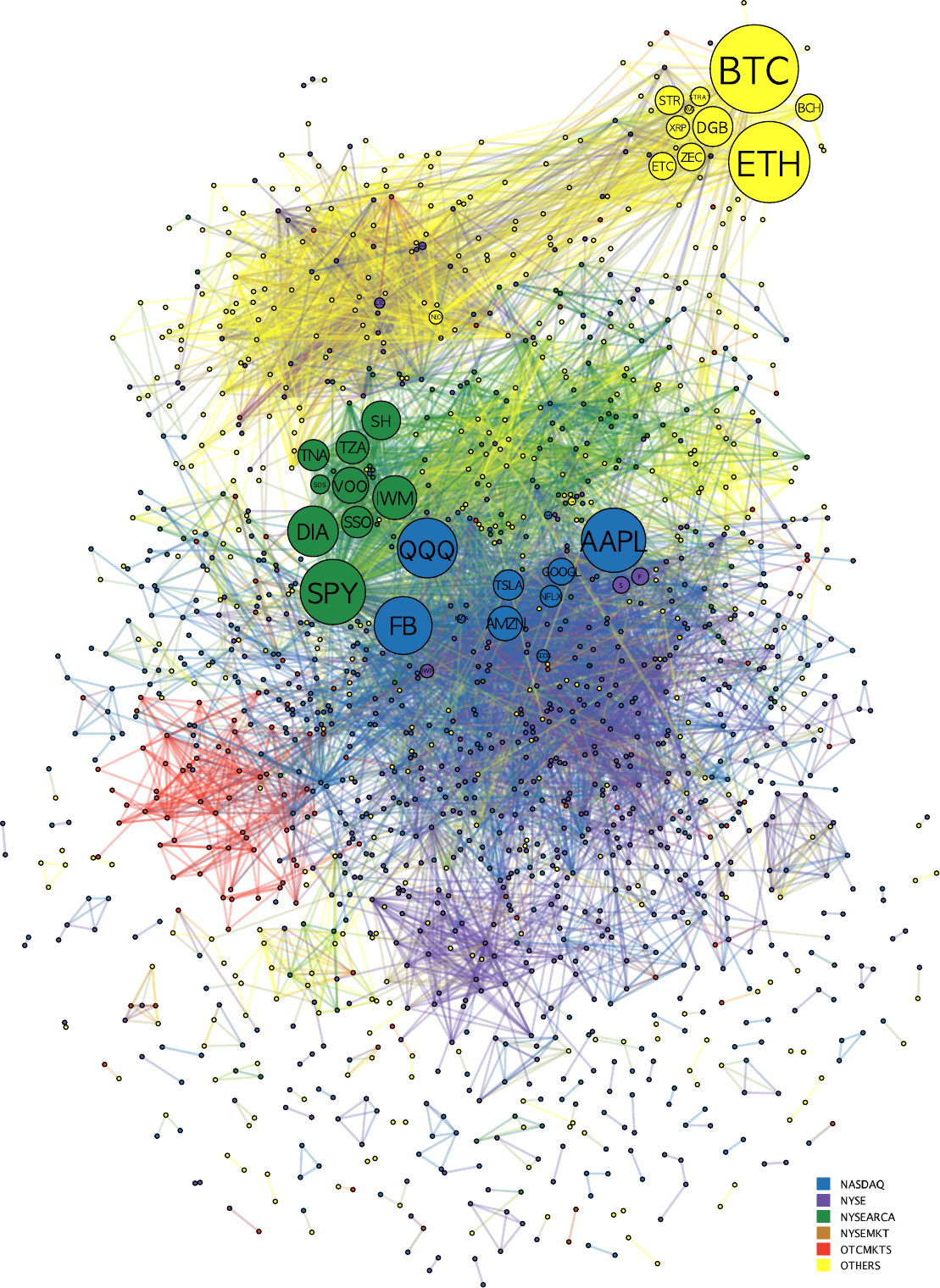}
    \caption{Co-occurrence graph of stocks mentioned in all tweets of our dataset. \texttt{OTCMKTS} stocks (red-colored) hold a peripheral position in the graph.     \label{fig:co-occurrences-cashtags-all-dataset}}
\end{figure*}

\subsection{Visualizing co-occurring stocks}
\label{sec:spam-graphs}
\begin{figure*}[t]
    \centering
    \includegraphics[width=0.85\textwidth]{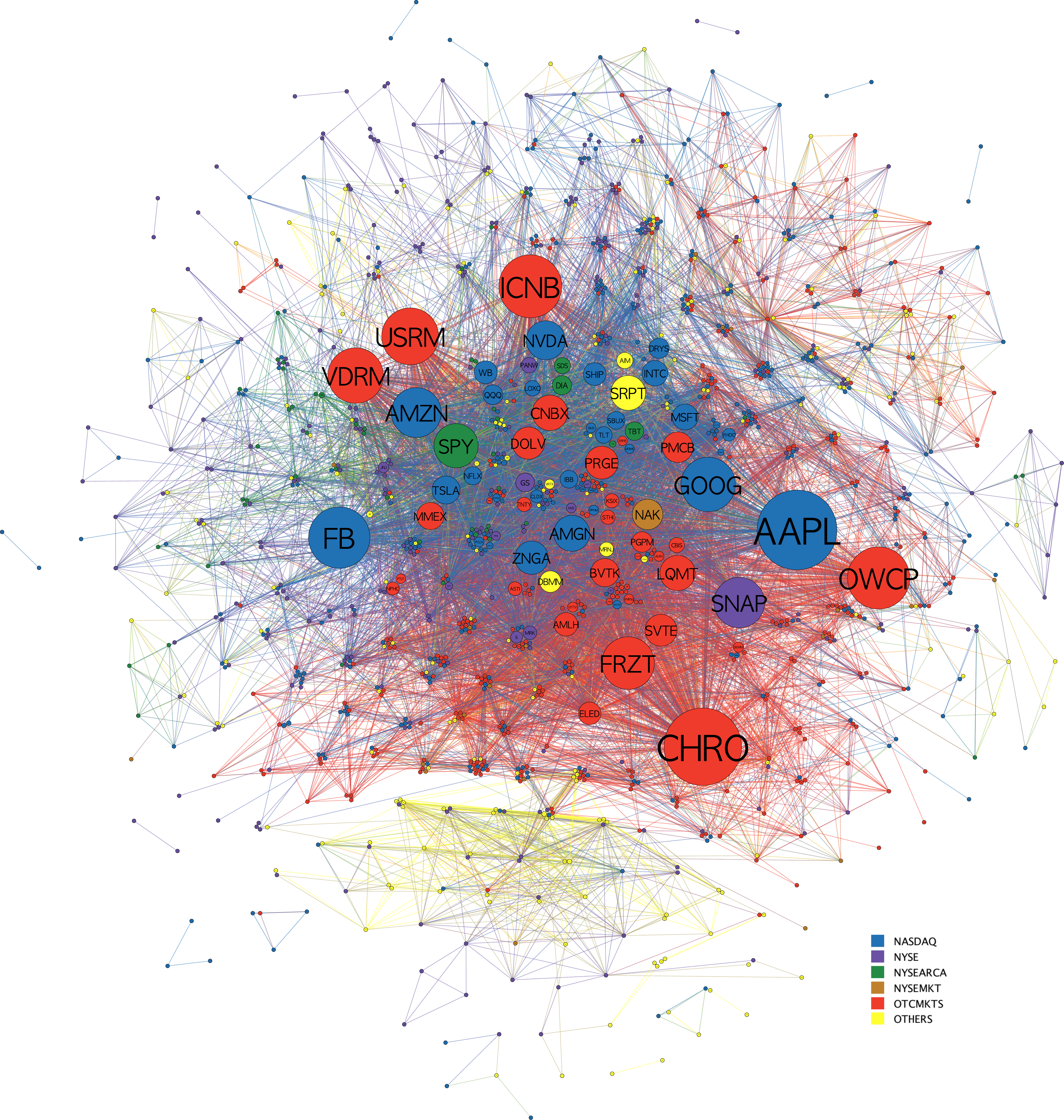}
    \caption{Co-occurrence graph of stocks mentioned in peak tweets. \texttt{OTCMKTS} stocks (red-colored) are central to the graph and are strongly connected with a number of \texttt{NASDAQ} stocks (blue-colored).     \label{fig:co-occurrences-cashtags-in-peaks}}
\end{figure*}

We begin by computing and visualizing the graph of co-occurring stocks for our whole dataset, and by comparing it with the graph of stocks that co-occur only in peak tweets. Our co-occurrence graphs represent the collective interconnections of stocks based on their paired presence within tweets. For the sake of clarity, graphs in figures~\ref{fig:co-occurrences-cashtags-all-dataset} and~\ref{fig:co-occurrences-cashtags-in-peaks} only show stocks whose degree $\ge 95$.

Figure~\ref{fig:co-occurrences-cashtags-all-dataset} shows the co-occurrence graph of stocks mentioned in all tweets of our dataset. Stocks are colored according to their market. As shown, the core of the graph is mainly composed of stocks belonging to \texttt{NASDAQ} (blue-colored) and \texttt{NYSEARCA} (green-colored) markets. In addition to stocks of the 5 markets already introduced, Figure~\ref{fig:co-occurrences-cashtags-all-dataset} also shows cashtags related to cryptocurrencies (yellow-colored). This is because in Twitter cryptocurrencies are labeled with cashtags, similarly to stocks. However, cryptocurrencies are not traded in regulated financial markets and hence in figure they are labeled as \texttt{OTHERS}. As shown, cryptocurrencies represent a large cluster of our graph, with a few highly important nodes such as \textit{Bitcoin} (\texttt{\$BTC}) and \textit{Ethereum} (\texttt{\$ETH}). Quite intuitively, cryptocurrencies are however well separated from the rest of the graph, meaning that they rarely co-occur in tweets with stocks traded in financial markets. Finally, in Figure~\ref{fig:co-occurrences-cashtags-all-dataset} \texttt{OTCMKTS} stocks (red-colored) cover only a small and peripheral portion of the graph.

In Figure~\ref{fig:co-occurrences-cashtags-in-peaks} we recreate the co-occurrence graph by only considering peak tweets. This time, the core of the graph is mainly composed of stocks from \texttt{NASDAQ} (blue-colored) and \texttt{OTCMKTS} (red-colored). More precisely, \texttt{OTCMKTS} stocks are not in the periphery of the graph, but instead are well interconnected with \texttt{NASDAQ} stocks. In addition, the degree of many \texttt{OTCMKTS} stocks is comparable to that of \texttt{NASDAQ} stocks. Intuitively, this means that \texttt{OTCMKTS} stocks appear very frequently in peak tweets, and that they often co-occur in such tweets with \texttt{NASDAQ} stocks.

\subsection{Analysis of co-occurring stocks by industrial classification}
\label{sec:spam-TRBC}

\begin{figure*}[t]
    \centering
    \begin{subfigure}[t]{.40\textwidth}
        \includegraphics[width=\textwidth]{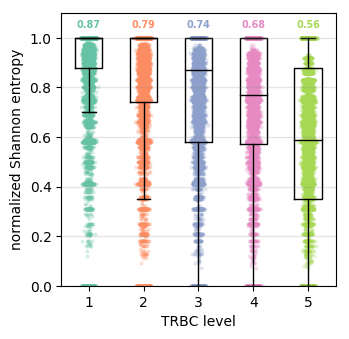}
        \caption{Peaks tweets.} \label{fig:tnorm-entropy-tweets-TRBC-peaks}
    \end{subfigure}\hspace{.05\textwidth}    \begin{subfigure}[t]{.40\textwidth}
        \includegraphics[width=\textwidth]{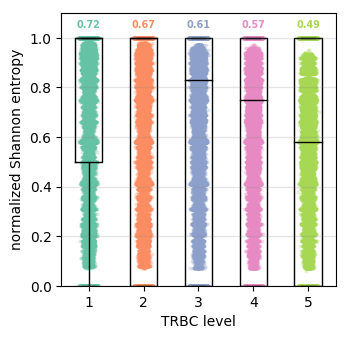}
        \caption{All dataset.}
        \label{fig:tnorm-entropy-tweets-TRBC-all-dataset}
    \end{subfigure}    \caption{Normalized Shannon entropy of the industrial (TRBC) classes of co-occurring stocks in tweets. TRBC level 1 has the finest grain, while level 5 has the coarsest grain. As shown, median entropy $> 0.5$ for all 5 TRBC levels, meaning that co-occurring companies in tweets are largely unrelated. Mean entropy values are reported above the boxplot and scatterplot distributions. Mean entropy measured for peak tweets is always higher than that measured for all tweets of the dataset. All differences are statistically significant.\label{fig:entropy-TRBC}}
\end{figure*}

Previous work have investigated the co-occurrences of stocks in weblogs and their relation to real-world events. In particular, authors of~\cite{kharratzadeh2012weblog} applied a clustering technique over a stock co-occurrences matrix, identifying a number of clusters containing highly correlated stocks. Results of this study highlighted that stocks that co-occur in blog articles as a consequence of real-world events, belong to the same industrial sector. In other words, results of~\cite{kharratzadeh2012weblog} support the assumption that stocks that legitimately appear related between one another in weblogs (or microblogs), are also related in real-world.
Thus, as a consequence of common sense and previous studies, it would be suspicious for some stocks to appear related (i.e., co-occurring) in microblogs, without being related (i.e., belonging to the same industrial sector) in real-world.

To evaluate whether co-occurring stocks in peak tweets of our dataset are also related in real-world, we exploited the TRBC classification previously introduced in Section~\ref{sec:dataset-financial}.
Specifically, for each tweet $t \in \mathbf{t}$ we measured the extent to which the stocks mentioned in $t$ belong to the same (or to different) TRBC class(es), for all the 5 hierarchical levels of TRBC. As a measurement for the difference in TRBC classes across stocks in a tweet, we leveraged the notion of \textit{entropy}. Thus, given a tweet $t \in \mathbf{t}$ containing $X$ distinct cashtags (i.e., each one associated to a different company) and the level $j$ of TRBC with $N_j$ classes, we first built the list of TRBC classes of the $X$ companies mentioned in $t$:
\[
\mathbf{c} = (c_1, c_2, \dots, c_X)
\]
Then, we computed the normalized Shannon entropy of the TRBC classes in $\mathbf{c}$, for TRBC level $j$, as:
\[
H^\mathbf{c}_{\text{norm}}(j) = \frac{-\sum\limits_{i=1}^{N_j}p^\mathbf{c}_i\log_{2}p^\mathbf{c}_i}{H_{\text{max}}(j)}
\]
where $p^\mathbf{c}_i$ is the empirical probability that TRBC class $i$ appears in $\mathbf{c}$, and $H_{\text{max}}(j)$ is the maximum theoretical entropy for TRBC level $j$:
\[
H_{\text{max}}(j) = -\log_{2}\frac{1}{X}
\]
Because of the normalization term, $0 \leq H^\mathbf{c}_{\text{norm}} \leq 1$. Thus, $H^\mathbf{c}_{\text{norm}} \sim 0$ implies companies of the same industrial sector, while $H^\mathbf{c}_{\text{norm}} \sim 1$ implies unrelated companies.

Intuitively, considering that the 5 TRBC levels are hierarchical, we expect $H^\mathbf{c}_{\text{norm}}$ to be higher (i.e., more heterogeneity) for fine-grained TRBC levels, while we expect $H^\mathbf{c}_{\text{norm}}$ to be lower (i.e., less heterogeneity) for the topmost, coarse-grained TRBC level. Results of this experiment, with TRBC level $j$ ranging from the lowest level $1$ to the topmost level $5$, are shown in Figure~\ref{fig:tnorm-entropy-tweets-TRBC-peaks}. For every TRBC level, a boxplot and a scatterplot show the distribution of normalized entropy measured for each peak tweet. As expected, $H^\mathbf{c}_{\text{norm}}$ actually lowers when considering coarse-grained TRBC levels, as shown by the median value of the boxplot distributions. Nonetheless, median $H^\mathbf{c}_{\text{norm}} > 0.5$ for all 5 TRBC levels, meaning that co-occurring companies in peak tweets are largely unrelated. Figure~\ref{fig:tnorm-entropy-tweets-TRBC-all-dataset} shows the result of the same measurement carried out on all tweets of our dataset, rather than only on peak tweets. Interestingly, the entropy measured in all our dataset is smaller than that measured for peak tweets, for all 5 TRBC levels. In turn, this means that co-occurring companies in peak tweets are overall less related than those co-occurring in tweets not belonging to a peak. Differences between the entropies measured for peak tweets and for all the dataset are statistically significant for all 5 TRBC levels, with all \textit{p}-values $< 0.01$ according to a 2-sample Kolmogorov-Smirnov test.

Notably, even for fine-grained TRBC levels, there is a minority of peak tweets for which we measured $H^\mathbf{c}_{\text{norm}} = 0$. These tweets might actually contain mentions to companies related also in real-world.

Summarizing, the results of this experiment seem to suggest that, overall, co-occurrences of stocks in peak tweets are not motivated by the fact that stocks belong to the same industrial or economic sectors.

\begin{comment}
\begin{figure}[h!]
\centering
\includegraphics[scale=0.3]{norm-entropy-tweets-TRBC-and-cashtag-number.png}
\caption{normalized entropy of tweets by TRBC and by cashtag number}
\label{fig:norm-entropy-tweets-TRBC-and-cashtag-number}
\end{figure}
\end{comment}

\subsection{Analysis of co-occurring stocks by market capitalization}
\label{sec:spam-market-cap}
Since real-world relatedness (as expressed by industrial classification) is not a plausible explanation for co-occurring stocks in our dataset, we now turn our attention to market capitalization. We are interested in evaluating whether a relation exists between the capitalization of co-occurring stocks. For instance, legitimate peak tweets could mention multiple stocks with similar capitalization. Conversely, malicious users could try to exploit the popularity of high-cap stocks by mentioning them together with low-cap ones.

\begin{figure*}[t]
    \centering
    \includegraphics[width=0.6\textwidth]{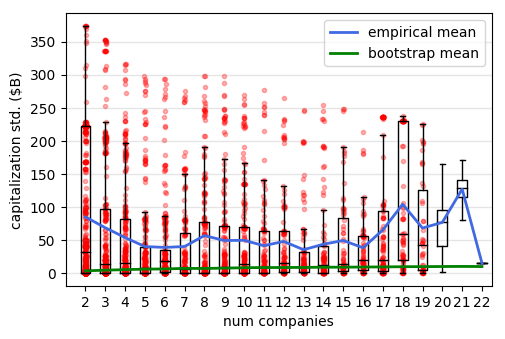}
    \caption{Standard deviation of the capitalization of co-occurring companies in peak tweets, and comparison with a bootstrap. The large measured standard deviation implies that high-cap companies co-occur with low-cap ones.\label{fig:cap-std-dev}}
\end{figure*}

One way to evaluate the similarity (or dissimilarity) in market capitalization of co-occurring stocks is by computing statistical measures of spread, \textit{standard deviation} (std.) being a straightforward one. Thus, for each peak tweet $t \in \mathbf{t}$ we computed the std. of the capitalization of all companies mentioned in $t$. Results are shown in Figure~\ref{fig:cap-std-dev}, where boxplots and scatterplots are depicted as a function of the number of distinct companies mentioned in tweets. Then, in order to understand whether the measured spread in capitalization is due to the intrinsic characteristics of our dataset (i.e., the underlying statistical distribution of capitalization) or to other factors, we compared mean values of our empirical measurements with the result of a \textit{bootstrap}. For bootstrapping the std. of tweets that mention $x$ companies, we randomly sampled $10,000$ groups of $x$ companies from our dataset. Then, for each of the $10,000$ random groups we computed the std. of the capitalization of the $x$ companies of the group. Finally, we averaged results over the $10,000$ groups. This procedure is executed for $x = 2, 3, \dots, 22$, thus covering the whole extent of Figure~\ref{fig:cap-std-dev}.

Results in figure highlight a large empirical std. between the capitalization of co-occurring companies. This means that in our peak tweets, high-cap companies co-occur with low-cap ones. Moreover, the measured std. is larger than that obtained with the bootstrap. In turn, this means that the large difference in capitalization can not be explained by the intrinsic characteristics of our dataset, but it is rather the consequence of an external action.

The previous experiment already lead to interesting results. However, it does not allow to draw insights into the possibly different characteristics of stocks traded in different markets. In order to evaluate the capitalization of co-occurring stocks, for stocks of different markets, we evaluated the \textit{assortativity} of the co-occurrence graph of stocks mentioned in peak tweets. The graph used for this experiment is the one depicted in Figure~\ref{fig:co-occurrences-cashtags-in-peaks}. The assortativity is computed on the capitalization of the nodes (i.e., companies) of the graph, rather than on their degree as it is typically done with this kind of analysis.

\begin{figure*}[t]
    \centering
    \begin{subfigure}[b]{.3\textwidth}
        \includegraphics[width=\textwidth]{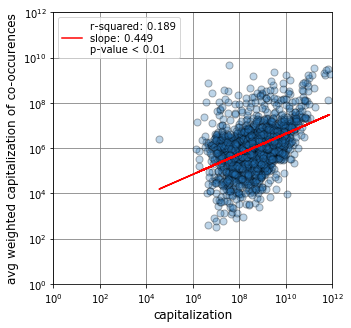}
        \caption{\texttt{NASDAQ}.\label{fig:assortativity-NASDAQ}}
    \end{subfigure}\hspace{.04\textwidth}    \begin{subfigure}[b]{.3\textwidth}
        \includegraphics[width=\textwidth]{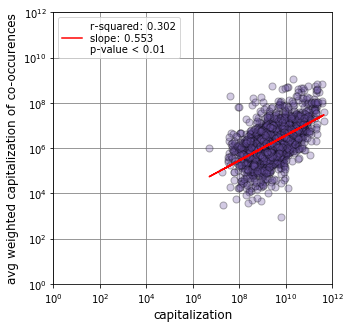}
        \caption{\texttt{NYSE}.\label{fig:assortativity-NYSE}}
    \end{subfigure}\hspace{.04\textwidth}    \begin{subfigure}[b]{.3\textwidth}
        \includegraphics[width=\textwidth]{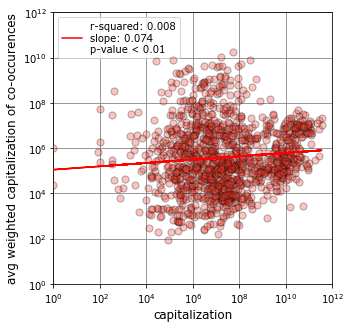}
        \caption{\texttt{OTCMKTS}.\label{fig:assortativity-OTCMKTS}}
    \end{subfigure}    \caption{Assortativity plots for the co-occurrence graph of stocks mentioned in peak tweets. Assortativity is computed out of stocks capitalization and results are grouped by market.\label{fig:assortativity}}
\end{figure*}

Specifically, for every stock, we compare its capitalization with the weighted mean of the capitalizations of its neighbors in the graph. The weighting factor is based on the number of co-occurrences between stocks (i.e., the weight of the edge in the co-occurrence graph). Results are presented in Figure~\ref{fig:assortativity} as scatterplots with a linear fit, and are grouped by market. In figure~\ref{fig:assortativity} we only show plots for \texttt{NASDAQ}, \texttt{NYSE}, and \texttt{OTCMKTS} since they represent the most interesting results. As shown, stocks of \texttt{NASDAQ} and \texttt{NYSE}, the most important markets of our dataset, are assortative (slopes equal $0.44$ and $0.55$). In other words, high-cap stocks of \texttt{NASDAQ} and \texttt{NYSE} typically co-occur with other high-cap stocks. This behavior is consistent with what one would intuitively expect. Conversely, \texttt{OTCMKTS} stocks are almost non-assortative at all, as demonstrated by slope $\sim 0$.

It is also important to note that while the assortativity of \texttt{NASDAQ} and \texttt{NYSE} stocks is higher when considering peak tweets instead of all the tweets of our dataset, for \texttt{OTCMKTS} stocks we measure the opposite behavior. This means that in peak tweets \texttt{OTCMKTS} stocks co-occur with high-cap stocks more often than when considering all our dataset.

\begin{comment}
\begin{figure*}[t]
    \centering
    \begin{subfigure}[b]{.3\textwidth}
        \includegraphics[width=\textwidth]{assortativity-dataset-NASDAQ.png}
        \caption{\texttt{NASDAQ}.\label{fig:assortativity-dataset-NASDAQ}}
    \end{subfigure}\hspace{.04\textwidth}    \begin{subfigure}[b]{.3\textwidth}
        \includegraphics[width=\textwidth]{assortativity-dataset-NYSE.png}
        \caption{\texttt{NYSE}.\label{fig:assortativity-dataset-NYSE}}
    \end{subfigure}\hspace{.04\textwidth}    \begin{subfigure}[b]{.3\textwidth}
        \includegraphics[width=\textwidth]{assortativity-dataset-OTCMKTS.png}
        \caption{\texttt{OTCMKTS}.\label{fig:assortativity-dataset-OTCMKTS}}
    \end{subfigure}    \caption{Assortativity plots for the co-occurrence graph of stocks mentioned in all tweets of the dataset. Assortativity is computed out of stocks capitalization and results are grouped by market.\label{fig:assortativity-dataset}}
\end{figure*}
\end{comment}

\subsection{Social and financial importance}
\label{sec:spam-importance}
So far, we demonstrated that tweets responsible for generating peaks, mention a large number of unrelated stocks, some of which are high-cap stocks while the others are low-cap ones. Adding to these findings, we are also interested in assessing the relation between the \textit{social} and \textit{financial importance} of our 30,032 stocks. Financial importance of a stock $i$ can be measured by its market capitalization $C_i$. Social importance can be quantified as the number of times a stock is mentioned in stock microblogs. Intuitively, we expect a positive correlation between stock capitalization and mentions, meaning that high-cap stocks are mentioned more frequently than low-cap stocks.
Notably, this positive relation has already been measured in a number of previous works, such as~\cite{mao2012correlating}, and has been leveraged for predicting stock prices.

\begin{comment}
\begin{table}[h]
	\footnotesize
	\centering
	\begin{tabular}{lcrrcrrcrr}
		\toprule
		&& \multicolumn{2}{c}{\textbf{pearson correlation}} && \multicolumn{2}{c}{\textbf{spearman's $\rho$}} && \multicolumn{2}{c}{\textbf{kendall's $\tau$}} \\
		\cmidrule{3-4} \cmidrule{6-7} \cmidrule{9-10}
		\textbf{markets} && \textit{all dataset} & \textit{peaks} && \textit{all dataset} & \textit{peaks} && \textit{all dataset} & \textit{peaks} \\
		\midrule
		\texttt{NASDAQ}	    && $0.7342$     & $0.9566$  && $0.4074$     & $0.0772$  && $0.2960$     & $0.0526$\\
		\texttt{NYSE}	    && $0.4197$     & $0.4534$  && $0.6347$     & $0.3497$  && $0.4703$     & $0.2452$\\
		\texttt{NYSEARCA}	&& $0.7070$     & $-0.3635$ && $0.4318$     & $0.1429$  && $0.2966$     & $0.1429$\\
		\texttt{NYSEMKT}	&& $0.0959$     & $0.0986$  && $0.1054$     & $0.0420$  && $0.0719$     & $0.0215$\\
		\texttt{OTCMKTS}	&& $-0.0108$	& $-0.0630$ && $0.0778$	    & $-0.2658$ && $0.0556$	    & $-0.1758$\\
		\bottomrule
	\end{tabular}
	\caption{Rank correlation between market capitalization and number of tweets.}
	\label{tab:market-tweets-correlation}
\end{table}
\end{comment}

\begin{table}[h]
	\footnotesize
	\centering
	\begin{tabular}{lcrrcrr}
		\toprule
		&& \multicolumn{2}{c}{\textbf{spearman's $\rho$}} && \multicolumn{2}{c}{\textbf{kendall's $\tau$}} \\
		\cmidrule{3-4} \cmidrule{6-7}
		\textbf{markets} && \textit{all dataset} & \textit{peaks} && \textit{all dataset} & \textit{peaks} \\
		\midrule
		\texttt{NASDAQ}	    && $0.4074$     & $0.0772$  && $0.2960$     & $0.0526$\\
		\texttt{NYSE}	    && $0.6347$     & $0.3497$  && $0.4703$     & $0.2452$\\
		\texttt{NYSEARCA}	&& $0.4318$     & $0.1429$  && $0.2966$     & $0.1429$\\
		\texttt{NYSEMKT}	&& $0.1054$     & $0.0420$  && $0.0719$     & $0.0215$\\
		\texttt{OTCMKTS}	&& $0.0778$	    & $-0.2658$ && $0.0556$	    & $-0.1758$\\
		\bottomrule
	\end{tabular}
	\caption{Rank correlation between market capitalization and number of tweets.}
	\label{tab:market-tweets-correlation}
\end{table}

By exploiting our data in Table~\ref{tab:dataset} we can make a first assessment of this relation over the whole dataset and compare it with that measured for peak tweets. Specifically, in Table~\ref{tab:market-tweets-correlation} we report the values of 2 well-known rank correlation measures -- namely, Spearman's rank correlation coefficient ($\rho$), and Kendall's rank correlation coefficient ($\tau$) -- between the capitalization of a stock and the number of tweets mentioning that stock. The rank correlation is computed for all stocks of the 5 markets. When considering all our dataset, for stocks of all markets, except \texttt{OTCMKTS}, we find a positive correlation confirming our previous hypothesis. Instead, stocks of \texttt{OTCMKTS} feature negligible rank correlations over all the dataset. Even more interestingly, the significant correlation measured over all the dataset for stocks of important markets almost completely disappears when only considering peak tweets. Furthermore, \texttt{OTCMKTS} stocks in peak tweets even feature a moderate negative correlation. In other words, results of this experiment imply that the less capitalized stocks in \texttt{OTCMKTS} are more likely to appear in peak tweets than the more capitalized ones, a behavior that is both counterintuitive and in contrast with results of previous works. In turn, this further highlights the presence of suspicious behaviors in peaks.

\begin{figure*}[t]
    \centering
    \includegraphics[width=\textwidth]{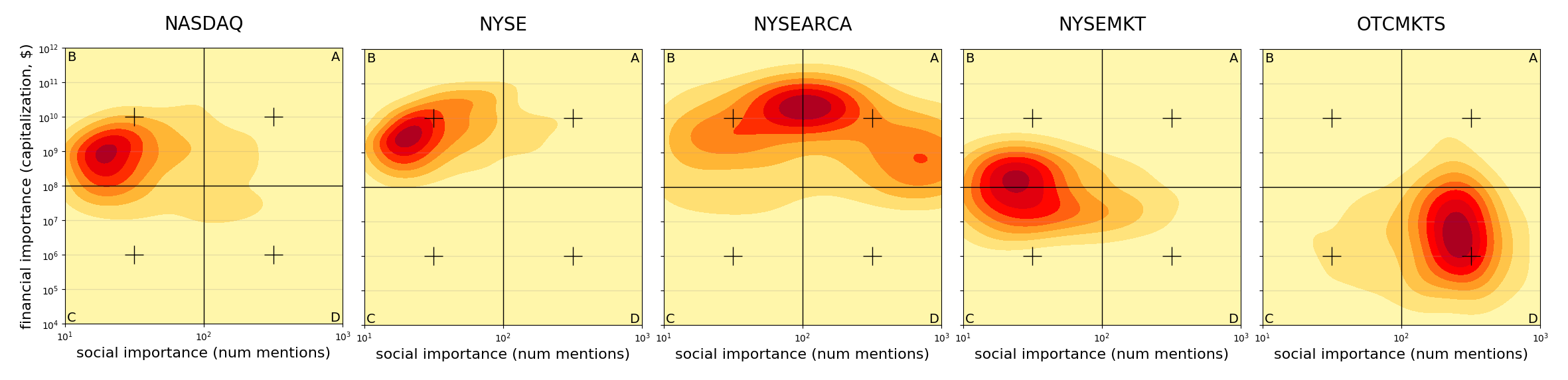}
    \caption{Kernel density estimation of social and financial importance, for stocks of the 5 considered markets. \texttt{OTCMKTS} stocks have a suspiciously high social importance despite their low financial importance, in contrast with stocks of all other markets.\label{fig:kde}}
\end{figure*}

\begin{comment}
    \begin{figure*}[t]
        \centering
        \begin{subfigure}[b]{.213\textwidth}
            \includegraphics[width=\textwidth]{kde-NASDAQ.png}
            \caption{\texttt{NASDAQ}.\label{fig:kde-NASDAQ}}
        \end{subfigure}        \begin{subfigure}[b]{.195\textwidth}
            \includegraphics[width=\textwidth]{kde-NYSE.png}
            \caption{\texttt{NYSE}.\label{fig:kde-NYSE}}
        \end{subfigure}        \begin{subfigure}[b]{.195\textwidth}
            \includegraphics[width=\textwidth]{kde-NYSEARCA.png}
            \caption{\texttt{NYSEARCA}.\label{fig:kde-NYSEARCA}}
        \end{subfigure}        \begin{subfigure}[b]{.195\textwidth}
            \includegraphics[width=\textwidth]{kde-NYSEMKT.png}
            \caption{\texttt{NYSEMKT}.\label{fig:kde-NYSEMKT}}
        \end{subfigure}        \begin{subfigure}[b]{.195\textwidth}
            \includegraphics[width=\textwidth]{kde-OTCMKTS.png}
            \caption{\texttt{OTCMKTS}.\label{fig:kde-OTCMKTS}}
        \end{subfigure}
        \caption{Kernel density estimation of social and financial importance, for stocks of the 5 considered markets. \texttt{OTCMKTS} stocks have a suspiciously high social importance despite their low financial importance, in contrast with stocks of all other markets.\label{fig:kde}}
    \end{figure*}
\end{comment}

With the goal of better evaluating the relationship between social and financial importance of stocks appearing in peaks, we also performed an additional experiment as follows. Given a stock $i$ and a peak $p$, we counted the number of times that $i$ is mentioned in peak tweets of $p$. We repeated this measurement for every peak $p$, and we computed the median value of these measurements that represents the social importance of stock $i$ in all peak tweets. Then, for every stock, we plotted its measurement of social importance versus that of financial importance, and we visually grouped stocks by their market. To avoid overplotting, we performed a bivariate (i.e., 2D) kernel density estimation, whose results are shown in Figure~\ref{fig:kde}. For the sake of clarity, we split the social--vs--financial space into 4 sectors. \textit{Sector A} (top-right) defines a region of space with stocks having both a high social and financial importance. Stocks in \textit{Sector B} (top-left) are characterized by high financial importance, but low social importance. Stocks in \textit{Sector C} (bottom-left) have both low social and financial importance, while stocks in \textit{Sector D} (bottom-right) have high social importance despite low financial importance.

By comparing stock densities of different markets in Figure~\ref{fig:kde}, we see that \texttt{OTCMKTS} stocks almost completely lay in \textit{Sector D}. All other markets have their stock densities mainly laying in \textit{Sector B} and \textit{Sector A}. In other words, \texttt{OTCMKTS} stocks have a suspiciously high social importance (i.e., they are mentioned in many tweets and across many peaks), despite their low financial importance. Results for all other markets are more intuitive, with \texttt{NYSEARCA} stocks achieving the best combination of social and financial importance. Summarizing, we measured a positive relation between social and financial importance when considering all stock microblogs shared during the 5 months of our study. However, when focusing our analysis on peaks in stock microblogs, we observed a suspicious behavior related to \texttt{OTCMKTS} stocks.

\makeatletter{}
\section{Analysis of suspicious users}
\label{sec:users}
In previous sections we identified a wide array of suspicious phenomena related to stock microblogs. In particular, peaks in microblog conversations about high-cap stocks are filled with mentions of low-cap (mainly \texttt{OTCMKTS}) stocks. Such mentions can not be explained by real-world stock relatedness. Moreover, the peaks in microblog conversations are largely caused by mass retweets. Despite not having been studied before, this scenario resembles those recently discovered when investigating the activities of bots tampering with social political discussions~\cite{ratkiewicz2011detecting,ferrara2016detection,cresci2017paradigm}. Unfortunately, systems for automatically detecting spam in stock microblogs are yet to be developed. However, recent scientific efforts lead to the development of several general-purpose bot and spam detection systems.

\subsection{Digital DNA for social bot detection}
\label{sec:users-ddna}
In this section we employ a state-of-the-art bot and spam detection system, specifically developed for spotting malicious group activities, to classify suspicious users~\cite{cresci2016dna,cresci2017tdsc}. The goal of this experiment is to assess whether users that shared/retweeted the suspicious peak tweets we previously identified, are classified as bots. In turn, this would bring definitive evidence of bot activities in the stock microblogs that we analyzed. The system in~\cite{cresci2016dna,cresci2017tdsc} performs bot detection in 2 steps. Firstly, it encodes the online behavior of a user into a string of characters that represents the \textit{digital DNA} of the user. Then, multiple digital DNA sequences, one for each user of the group under investigation, are compared between one another by means of string mining and bioinformatics algorithms. The system classifies as bots those users that have suspiciously high similarities among their digital DNA sequences. Notably, the system in~\cite{cresci2016dna,cresci2017tdsc} proved capable of accurately detecting also ``evolved'' bots ($F1 = 0.97$), such as those described in~\cite{ferrara2016rise}.

\begin{figure*}[t]
    \captionsetup[subfigure]{labelformat=empty}
    \centering
    \begin{subfigure}[t]{.25\textwidth}
        \includegraphics[width=\textwidth]{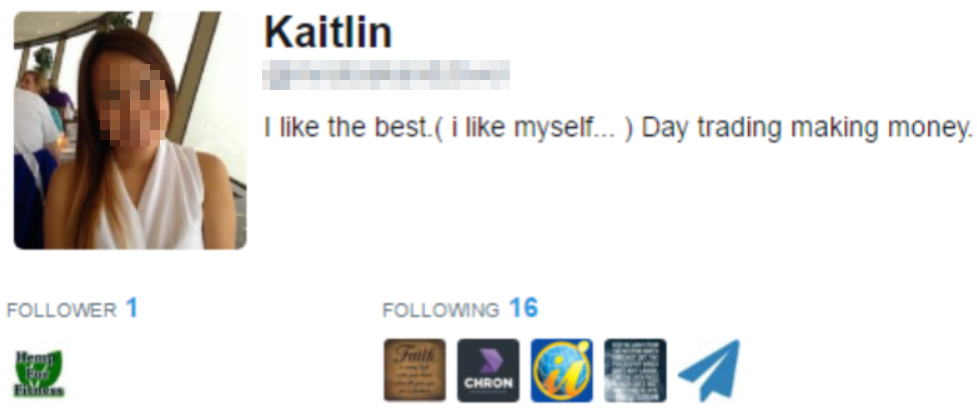}
    \end{subfigure}\hspace{.05\textwidth}    \begin{subfigure}[t]{.25\textwidth}
        \includegraphics[width=\textwidth]{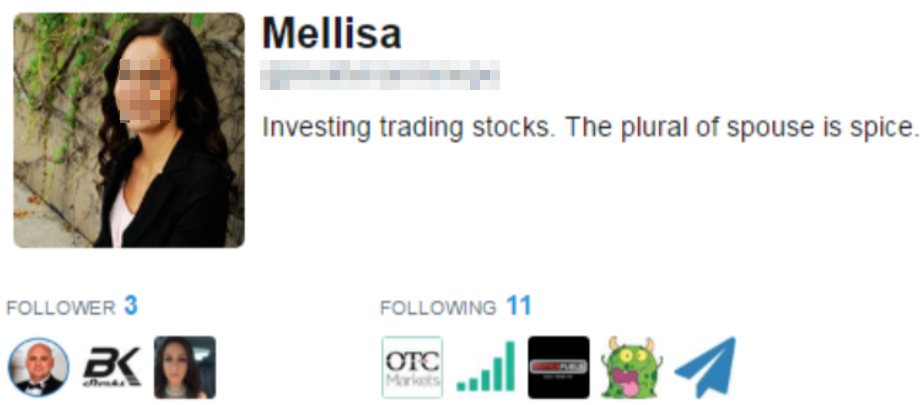}
    \end{subfigure}\hspace{.05\textwidth}    \begin{subfigure}[t]{.25\textwidth}
        \includegraphics[width=\textwidth]{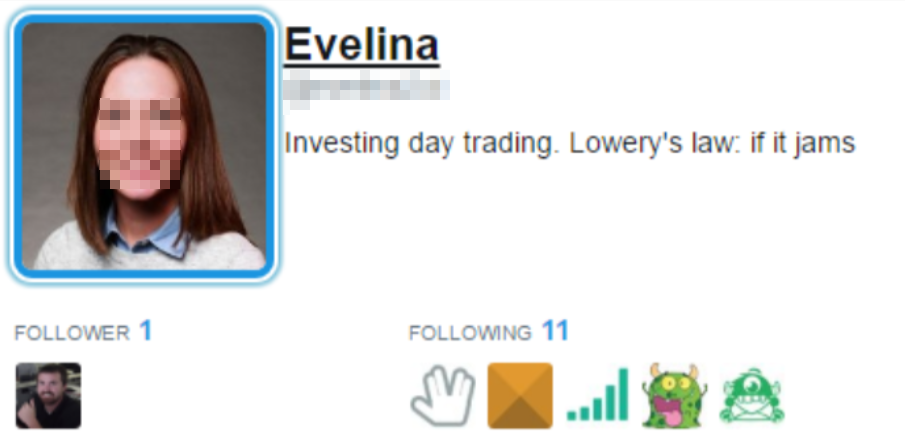}
    \end{subfigure}\\
    \begin{subfigure}[t]{.25\textwidth}
        \includegraphics[width=\textwidth]{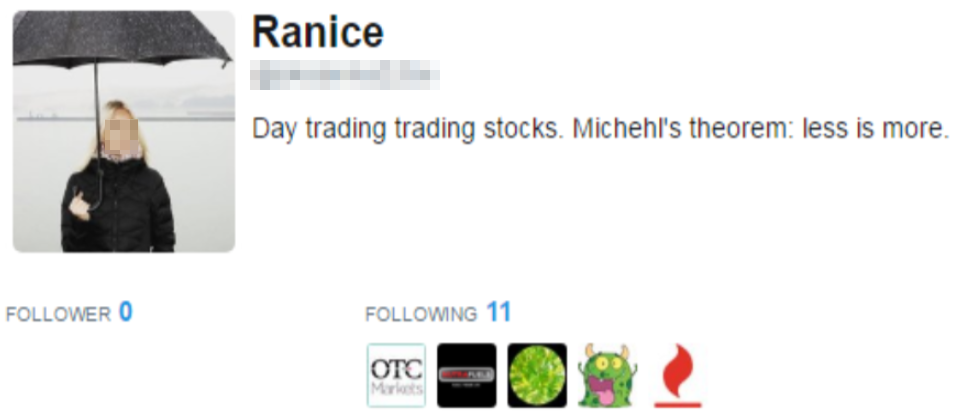}
    \end{subfigure}\hspace{.05\textwidth}    \begin{subfigure}[t]{.25\textwidth}
        \includegraphics[width=\textwidth]{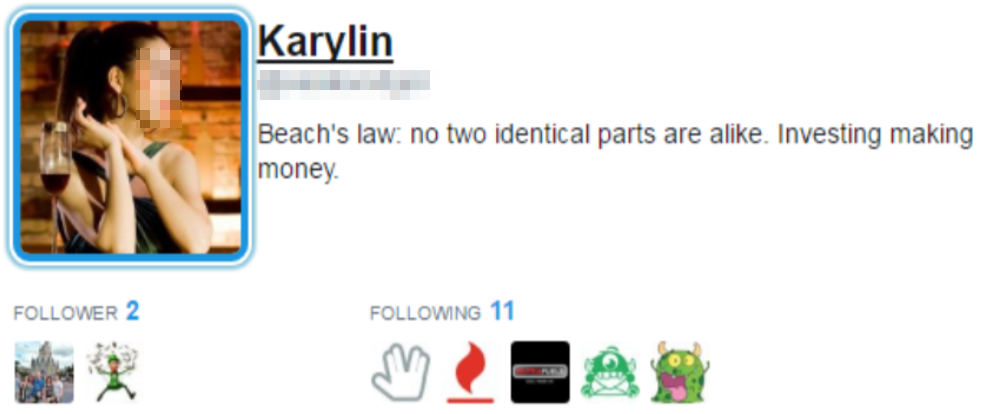}
    \end{subfigure}\hspace{.05\textwidth}    \begin{subfigure}[t]{.25\textwidth}
        \includegraphics[width=\textwidth]{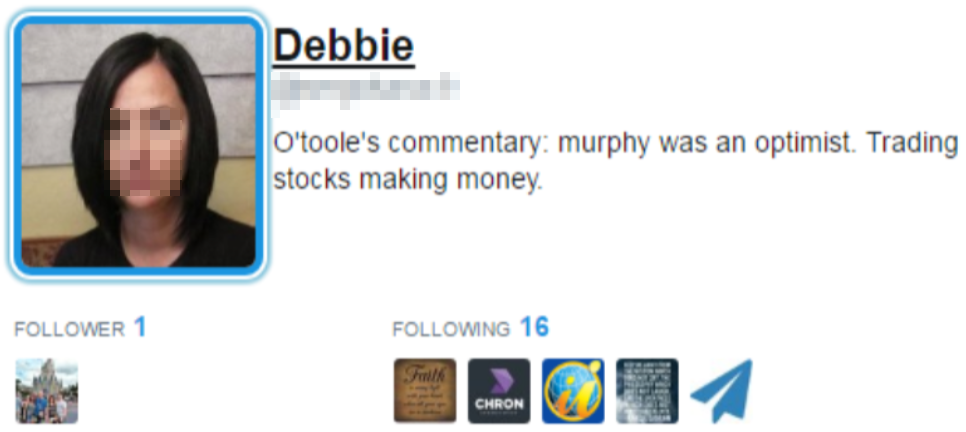}
    \end{subfigure}
    \caption{Examples of suspicious users classified as bots. The many characteristics shared between all these users (e.g., name, profile picture, social links) support the hypothesis that they are part of a larger botnet.\label{fig:bot-examples}}
\end{figure*}

\begin{figure*}[t]
    \captionsetup[subfigure]{labelformat=empty}
    \centering
    \begin{subfigure}[t]{.4\textwidth}
        \includegraphics[width=\textwidth]{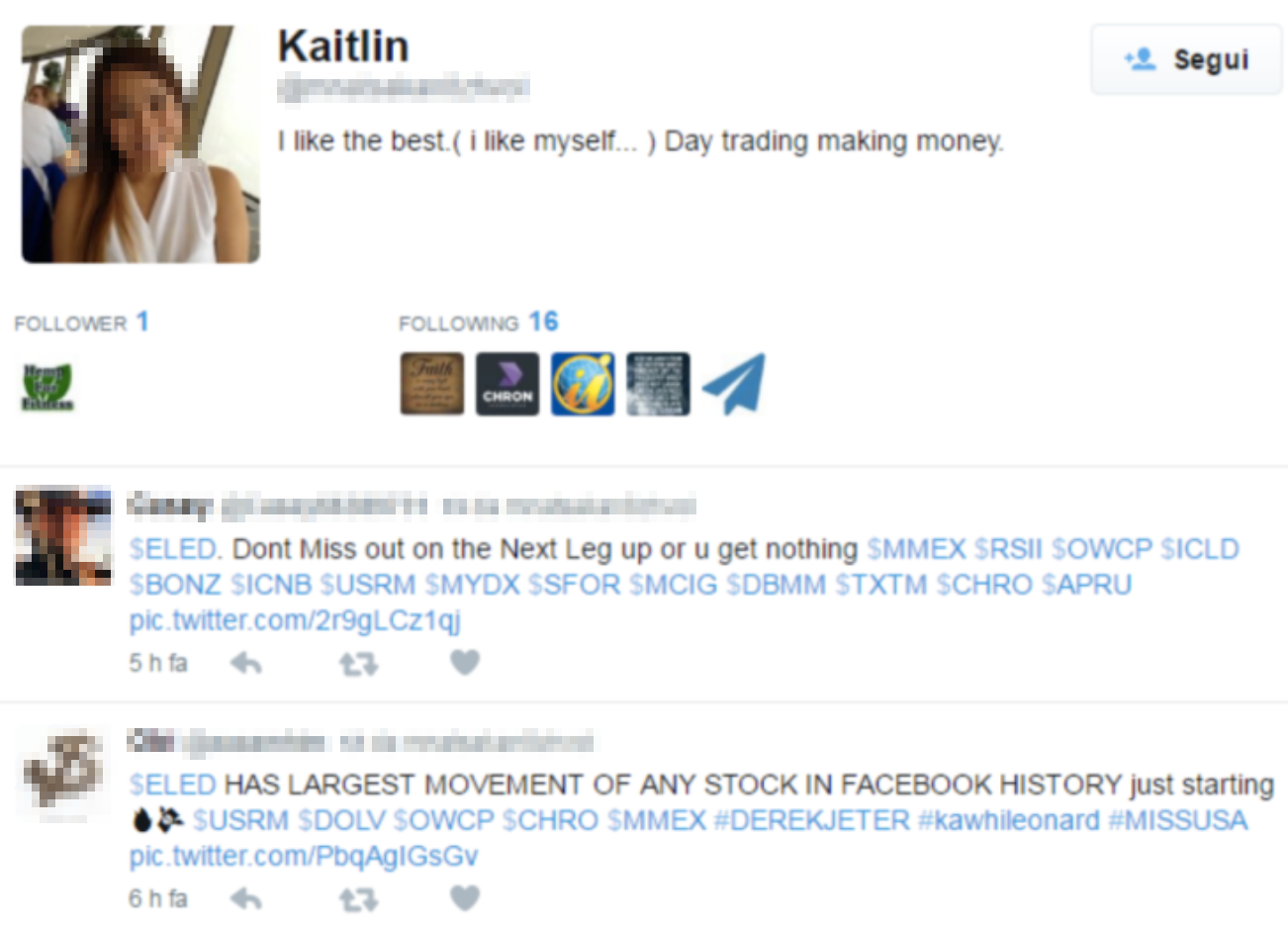}
    \end{subfigure}\hspace{.1\textwidth}    \begin{subfigure}[t]{.4\textwidth}
        \includegraphics[width=\textwidth]{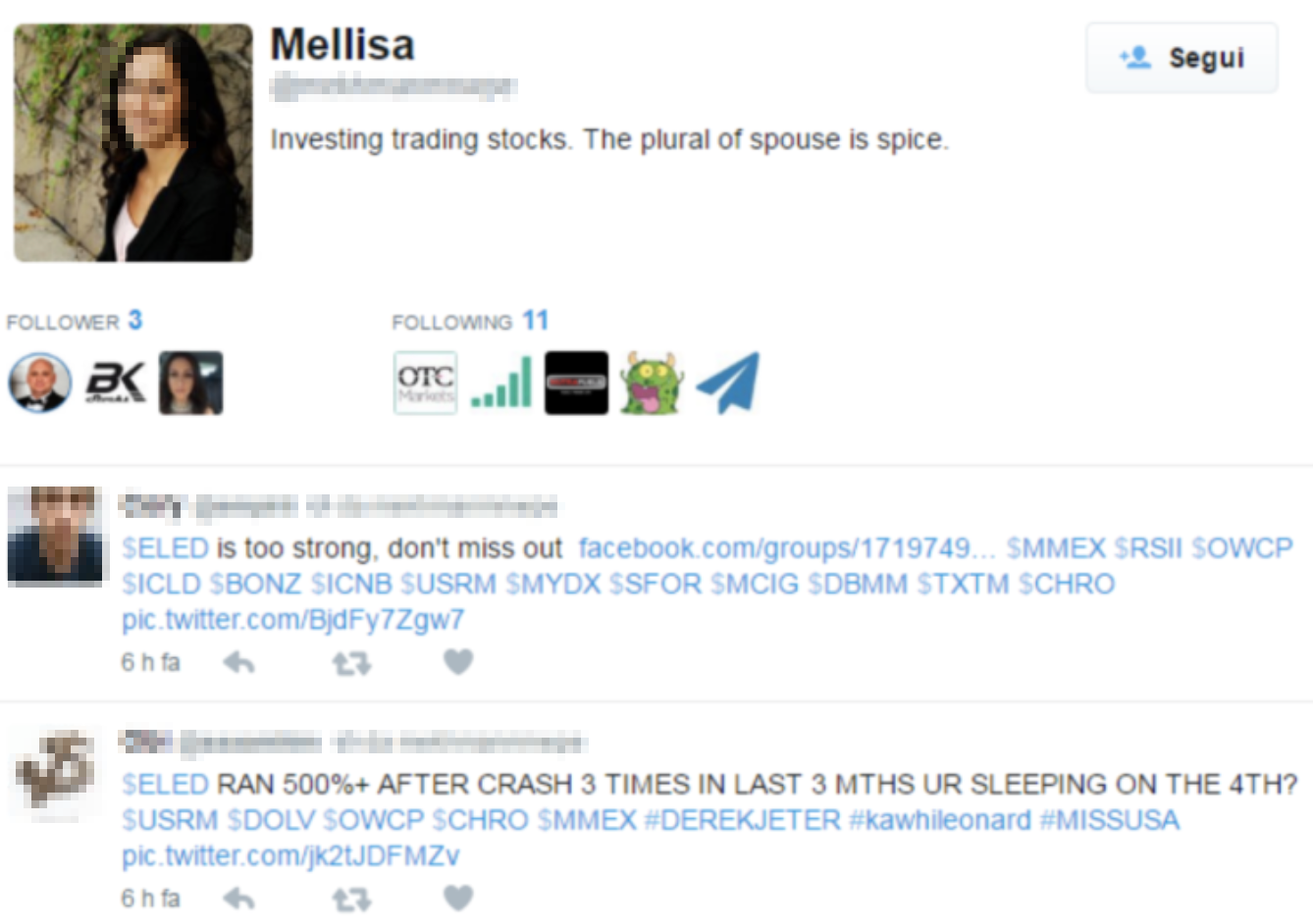}
    \end{subfigure}    \caption{Example tweets from two suspicious users classified as bots.\label{fig:spambots-tweets}}
\end{figure*}

Because of the computationally intensive analyses performed by~\cite{cresci2016dna,cresci2017tdsc}, we constrained this experiment to the 100 largest peaks (i.e., those generated by the greatest number of tweets) of our dataset. Starting from those top-100 peaks, we then analyzed the $25,957$ distinct users that shared or retweeted at least one peak tweet. Behavioral information needed by the detection system to perform user classification have been collected by crawling the Twitter timelines of such $25,957$ users. Notably, the bot detection system classified as much as $18,509$ (71\%) of the analyzed users as bots. Figure~\ref{fig:bot-examples} shows 6 examples of users classified as bots, while Figure~\ref{fig:spambots-tweets} shows some tweets of the same users. A manual analysis of a subset of bots allowed to identify characteristics shared between all the users (e.g., similar name, join date, profile picture, etc.), supporting the hypothesis that they are part of a larger botnet. Users classified as bots also feature very high retweet rates (ratio of retweets over all posted tweets), thus explaining the large number of retweets in our peaks and among \texttt{OTCMKTS} stock microblogs.

We obtained these results by analyzing only the 100 largest detected peaks, therefore analyses of minor peaks might yield different results. Nonetheless, the overwhelming ratio of bots that we discovered among large peaks discussing popular stocks, raises serious concerns over the reliability of stock microblogs.

\subsection{Twitter bot detection}
\label{sec:users-twitter}
In previous experiment we relied on a state-of-the-art bot detection technique in order to classify our accounts. Here, following a procedure originally used in~\cite{cresci2017paradigm}, we also evaluate whether Twitter itself detected and suspended the suspicious accounts that we identified. In fact, accounts that are suspected to perform malicious activities or that violate Twitter's terms of service, get suspended by Twitter.

To carry out this experiment we exploited Twitter's responses to API calls and, in particular, Twitter error codes\footnote{\url{https://developer.twitter.com/en/docs/basics/response-codes.html}}. Given a query to a specific account, Twitter APIs reply with information regarding the status of the queried account. API queries to a suspended account result in Twitter issuing \textit{error code 63}. Instead, for accounts that are still active, Twitter replies with the full metadata information of the account, without issuing any error.

Results of this experiment show that, out of the $25,957$ suspicious accounts, as much as $9,490$ (37\%) accounts have been suspended by Twitter somewhen between November 2017 and May 2018. This result is a clear demonstration that many of the accounts responsible for creating the peaks in financial discussions, are actually bots. It is not surprising that the \textit{digital DNA}-based technique~\cite{cresci2016dna,cresci2017tdsc} detected more bots than Twitter ($18,509$ versus $9,490$). Indeed, it has been recently demonstrated that state-of-the-art detection techniques are more effective than Twitter at detecting sophisticated bots~\cite{cresci2017paradigm}. Moreover, to avoid closing accounts of legitimate users by mistake, Twitter is typically conservative with its suspension policy. Finally, there is a very large overlap between the accounts suspended by Twitter and those labeled as bots via the digital DNA technique: $8,887$ out of $9,490$ accounts ($\sim 94\%$ of all Twitter suspensions).

\makeatletter{}
\section{Discussion}
\label{sec:discussion}
Results of our extensive investigation highlighted the presence of spam and bot activity in stock microblogs. For the first time, we described an advertising practice that we called \textit{cashtag piggybacking}, where many financially unimportant (low-cap) stocks are massively mentioned in microblogs together with a few financially important (high-cap) stocks. Analyses of suspicious users suggest that the advertising practice is carried out by large groups of coordinated social bots. Considering the already demonstrated relation between social and financial importance~\cite{mao2012correlating}, a possible outcome expected by perpetrators of this advertising practice is the increase in financial importance of the low-cap stocks, by exploiting the popularity of high-cap ones.

The potential negative consequences of this new form of financial spam are manifold. On the one hand, unaware investors (e.g., noise traders) could be lured into believing that the social importance of promoted stocks have a basis in reality. On the other hand, also the multitude of automatic trading systems that feed on social information, could be tricked into buying low value stocks. Market collapses such as the \textit{Flash Crash}, or disastrous investments such as that of \textit{Cynk Technology}, could occur again in the future, with dire consequences. For this reason, a favorable research avenue for the future involves quantifying the impact of social bots and microblog financial spam in stock prices fluctuations, similarly to what has already been done at the dawn of financial e-mail spam.

To the best of our knowledge, this is the first exploratory study on the presence of spam and bot activity in stock microblogs. As such, future works related to the characterization and detection of financial spam in microblogs, are much desirable. Indeed, no automatic system for the detection of financial spam in microblogs has been developed to date. To overcome this limitation, in our analyses we employed a general-purpose bot detection system. However, such approach hardly scales on the massive number of users, both legitimate and automated, involved in financial discussions on microblogs. Hence, another promising direction of research involves with the development of tools and techniques for promptly detecting promoted stocks, thus avoiding the need for a cumbersome user classification. In addition, a strict characterization of the social bots involved in \textit{cashtag piggybacking} spam campaigns (e.g., their behavior and network characteristics), is also needed.

Finally, we believe it is useful -- and worrying at the same time -- to demonstrate the presence of bot activity in stock microblogs. Finance thus adds to the growing list of domains recently tampered by social bots -- joining the political, social, and commercial domains, to name but a few.

\makeatletter{}
\section{Conclusions}
\label{sec:conc}
Motivated by the widespread presence of social bots, we carried out the first large-scale, systematic analysis on the presence and impact of spam and bot activity in stock microblogs. By cross-checking 9M stock microblogs from Twitter with financial information from Google Finance, we uncovered a malicious practice aimed at promoting low-value stocks by exploiting the popularity of high-value ones. In these so-called \textit{cashtag piggybacking} spam campaigns, many stocks with low market capitalization, mainly traded in \texttt{OTCMKTS}, are mentioned in microblogs together with a few high capitalization stocks traded in \texttt{NASDAQ} and \texttt{NYSE}. We showed that such co-occurring stocks are not related by economic and industrial sector. Moreover, the large discussion spikes about low-value stocks are due to mass, synchronized retweets. Finally, an analysis of retweeting users classified 71\% of them as bots, and 37\% of them were subsequently suspended by Twitter.

Given the severe consequences that this new form of financial spam could have on unaware investors as well as on automatic trading systems, our results call for the prompt adoption of spam and bot detection techniques in all applications and systems that exploit stock microblogs.

\begin{acks}

This research is supported in part by the \grantsponsor{001}{EU H2020 Program INFRAIA-1-2014-2015: Research Infrastructures}{} under Grant No.: \grantnum{001}{654024} \textit{SoBigData: Social Mining \& Big Data Ecosystem}.
\end{acks}

\bibliographystyle{ACM-Reference-Format}
\bibliography{manuscript}

\newpage
\makeatletter{}\section{Supplementary materials}
\label{sec:supplement}

\begin{printonly}
Supplementary materials are available in the online version of this paper.
\end{printonly}
 
\begin{screenonly}

\begin{figure}[t]
\centering
\includegraphics[width=0.375\textwidth]{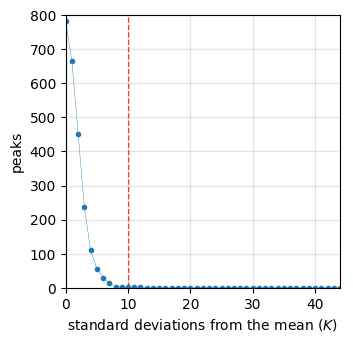}
\caption{Number of peaks detected, as a function of $K$ for \texttt{OTCMKTS} stocks.\label{fig:peaks-as-function-of-k-otc-tweets}}
\end{figure}

\begin{figure*}[t]
    \centering
    \captionsetup{justification=centering}
    \begin{subfigure}[b]{.24\textwidth}
        \includegraphics[width=\textwidth]{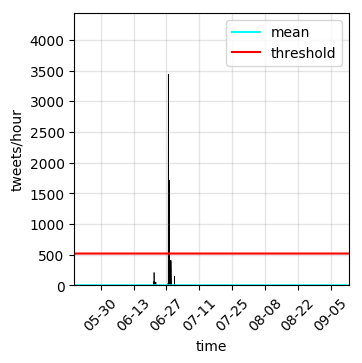}
        \caption{\texttt{\$UPZS} (\textit{Unique Pizza \& Subs Corp.}).\label{fig:timeserie-UPZS}}
    \end{subfigure}    \begin{subfigure}[b]{.24\textwidth}
        \includegraphics[width=\textwidth]{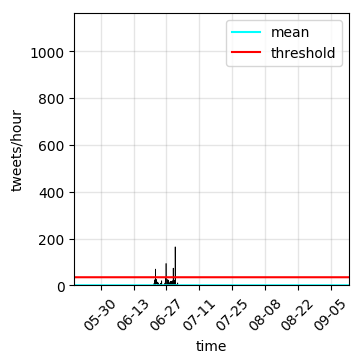}
        \caption{\texttt{\$KNSC} (\textit{Kenergy Scientific, Inc.}).\label{fig:timeserie-KNSC}}
    \end{subfigure}    \begin{subfigure}[b]{.24\textwidth}
        \includegraphics[width=\textwidth]{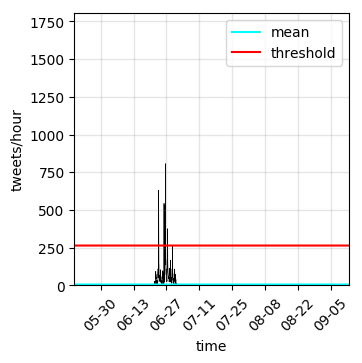}
        \caption{\texttt{\$INNV} (\textit{Innovus Pharmaceuticals, Inc.}).\label{fig:timeserie-INNV}}
    \end{subfigure}    \begin{subfigure}[b]{.24\textwidth}
        \includegraphics[width=\textwidth]{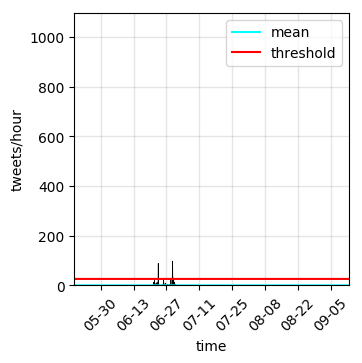}
        \caption{\texttt{\$NNSR} (\textit{NanoSensors, Inc.}).\label{fig:timeserie-NNSR}}
    \end{subfigure}    \caption{Examples of stock time series, for 4 \texttt{OTCMKTS} stocks. Mean values are marked with cyan solid lines and thresholds above which peaks are detected ($K = 7$) are marked with red solid lines.\label{fig:time-series-otcmkts}}
\end{figure*}

The data collection for this study is based on an official list of 6,689 high-cap stocks released by \texttt{NASDAQ}, as detailed in Section~\ref{sec:dataset}. As a consequence, we have complete data (e.g., the full time series) only for the 6,689 stocks of the list. Conversely, we have data about \texttt{OTCMKTS} stocks only if and when they co-occur with stocks of the list.

Because of this limitation in our dataset, some of our results regarding \texttt{OTCMKTS} stocks could be biased. In order to rule out this possibility, we carried out an additional data collection phase. Specifically, we collected all tweets shared between June 21 and July 2 2017 (i.e., 12 days), containing at least one cashtag of an \texttt{OTCMKTS} stock. This data collection lasted only for 12 days because of the large number (i.e., 22,956) of cashtags that we had to monitor, each one corresponding to a search keyword of a Twitter Streaming crawler. This new dataset is not biased towards high-cap stocks and represents a clear and complete picture of Twitter discussions about \texttt{OTCMKTS} stocks. Of all the tweets collected that contain at least one cashtag OTCMKTS, 51\% of them contain exactly one cashtag OTCMKTS.

We analyzed this dataset in the same way as the one obtained from the \texttt{NASDAQ} list. Specifically, we carried out the steps described in Section~\ref{sec:peaks}. Figure~\ref{fig:peaks-as-function-of-k-otc-tweets} shows the number of peaks detected in the dataset about \texttt{OTCMKTS} stocks, as a function of the parameter $K$ -- that is, the number of standard deviations from the mean needed to detect a peak in a stock time series. As shown in figure, by adopting the threshold chosen in Section~\ref{sec:peaks} ($K = 10$) we obtain no peaks at all. This is in sharp contrast with the result shown in Figure~\ref{fig:peaks-as-function-of-k}, obtained by analyzing the dataset derived from the \texttt{NASDAQ} list. Indeed, the overall volume of tweets that mention only \texttt{OTCMKTS} stocks is very low. In other words, this means that almost no tweet at all only mentions \texttt{OTCMKTS} stocks, while instead \texttt{OTCMKTS} stocks are almost only mentioned together with some other high-cap stock. This result obtained by analyzing the complete dataset of \texttt{OTCMKTS} stocks further supports our findings reported in Section~\ref{sec:piggyback}.

\begin{figure*}[t]
    \captionsetup[subfigure]{labelformat=empty}
    \centering
    \begin{subfigure}[t]{.4\textwidth}
        \includegraphics[width=\textwidth]{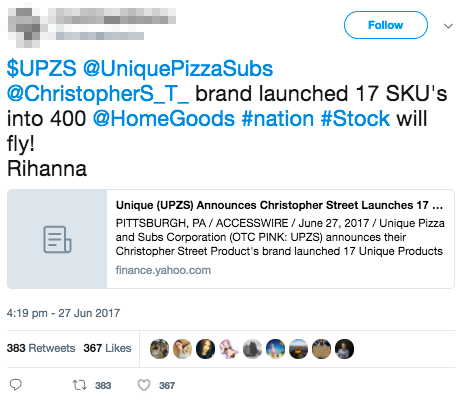}
    \end{subfigure}\hspace{.1\textwidth}    \begin{subfigure}[t]{.4\textwidth}
        \includegraphics[width=\textwidth]{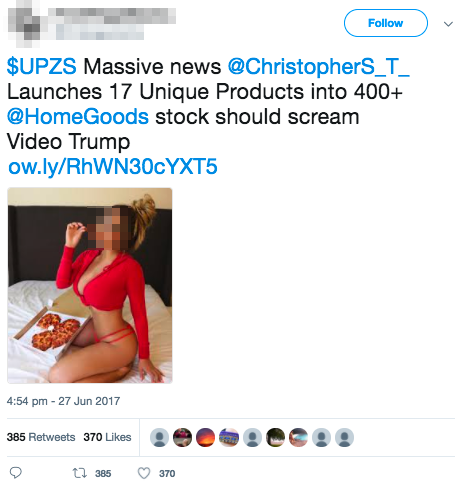}
    \end{subfigure}    \caption{Examples of tweets with just one \texttt{OTCMKTS} cashtag.\label{fig:one-OTC-tweets}}
\end{figure*}

If we lower the threshold needed to detect a peak in a stock time series from $K = 10$ to $K = 7$, we end up with 14 peaks, as shown in the examples of Figure~\ref{fig:time-series-otcmkts}. Such peaks are generally way lower than those measured for stocks of the \texttt{NASDAQ} list, which are shown in Figure~\ref{fig:time-series}. The 14 peaks detected in the \texttt{OTCMKTS} stocks dataset are still largely caused by mass retweets. Interestingly, Figure~\ref{fig:one-OTC-tweets} shows two examples of such peak tweets. As shown in figure, the tweets contain only the \texttt{\$UPZS} cashtag, related to the \textit{Unique Pizza \& Subs Corp.} company. Despite not showing the \textit{cashtag piggybacking} behavior, it is clear that these tweets are still aimed at exploiting some highly popular topics, in order to publicize the \texttt{\$UPZS} stock. Indeed, the left-hand side tweet of Figure~\ref{fig:one-OTC-tweets} ends with the keyword ``Rihanna'', while the right-hand side one ends with a link to a video related to (Donald) Trump and a provocative picture. Considering that there is clearly no relation between the content of the tweets in Figure~\ref{fig:one-OTC-tweets} and the pop singer Rihanna or the US President Donald Trump, this is just another way to \textit{piggyback} a stock on top of a popular discussion topic.

Summarizing, an analysis of the \texttt{OTCMKTS} stocks dataset described in this section shows that \texttt{OTCMKTS} stocks are almost only tweeted in conjunction with other high-cap stocks. This result supports our previous findings. Furthermore, we also uncovered a minority of  \texttt{OTCMKTS} stocks that do not feature the \textit{cashtag piggybacking} behavior, but rather that piggyback stocks on top of trending keywords.

\end{screenonly} 

\end{document}